\documentclass[conference,letterpaper]{IEEEtran}

\usepackage[utf8]{inputenc}
\usepackage[T1]{fontenc}
\usepackage{url}
\usepackage{ifthen}
\usepackage{cite}
\usepackage[cmex10]{amsmath} 

\IEEEoverridecommandlockouts
\newtheorem{Theorem}{Theorem}
\newtheorem{Lemma}{Lemma}
\newtheorem{prop}{Proposition}
\newtheorem{Corollary}[Theorem]{Corollary}
\newtheorem{defi}{Definition}

\usepackage{algorithm}
\usepackage{algorithmic}
\usepackage{graphicx}
\usepackage{textcomp}
\usepackage{stfloats}
\usepackage{subfigure}
\usepackage{amssymb,amsmath,bm}
\usepackage{cite}
\usepackage{times}
\usepackage{latexsym}
\usepackage{array}
\usepackage{fancyhdr}
\usepackage{psfrag}
\usepackage{multirow}
\usepackage{xcolor}
\usepackage{color}
\usepackage{makecell}
\usepackage{threeparttable}
\usepackage{hyperref}
\usepackage[OT1]{fontenc}
\allowdisplaybreaks[1]

\begin{document}
\title{Unsourced Random Access in MIMO Quasi-Static Rayleigh Fading Channels with Finite Blocklength}



 \author{%
   \IEEEauthorblockN{Junyuan Gao\IEEEauthorrefmark{1},
                     Yongpeng Wu\IEEEauthorrefmark{1},
                     Giuseppe Caire\IEEEauthorrefmark{2},
                     Wei Yang\IEEEauthorrefmark{3},
                     and Wenjun Zhang\IEEEauthorrefmark{1}}
   \IEEEauthorblockA{\IEEEauthorrefmark{1}%
                    Department of Electronic Engineering, Shanghai Jiao Tong University, Shanghai 200240, China, \\
                    sunflower0515@alumni.sjtu.edu.cn, \{yongpeng.wu, zhangwenjun\}@sjtu.edu.cn}
   \IEEEauthorblockA{\IEEEauthorrefmark{2}%
                    Communications and Information Theory Group, Technische Universit{\"a}t Berlin, Berlin 10587, Germany, caire@tu-berlin.de}
   \IEEEauthorblockA{\IEEEauthorrefmark{3}%
                    Qualcomm Technologies, Inc., San Diego, CA 92121, USA, weiyang@qti.qualcomm.com}
 }

\maketitle

\begin{abstract}
   This paper explores the fundamental limits of unsourced random access (URA) with a random and unknown number ${\rm{K}}_a$ of active users in MIMO quasi-static Rayleigh fading channels. First, we derive an upper bound on the probability of incorrectly estimating the number of active users. We prove that it exponentially decays with the number of receive antennas and eventually vanishes, whereas reaches a plateau as the power and blocklength increase. Then, we derive non-asymptotic achievability and converse bounds on the minimum energy-per-bit required by each active user to reliably transmit $J$ bits with blocklength $n$. Numerical results verify the tightness of our bounds, suggesting that they provide benchmarks to evaluate existing schemes. The extra required energy-per-bit due to the uncertainty of the number of active users decreases as $\mathbb{E}[{\rm{K}}_a]$ increases. Compared to random access with individual codebooks, the URA paradigm achieves higher spectral and energy efficiency. Moreover, using codewords distributed on a sphere is shown to outperform the Gaussian random coding scheme in the non-asymptotic regime.
\end{abstract}

\section{Introduction} \label{section1}

  Massive random access is expected to be one of the main use cases in future wireless networks, in which there are a large number of potential users and a fraction of them are active at any given time~\cite{wuyp}. Supporting reliable transmission of short packets for active users with limited channel uses and stringent energy constraints is critically required.
  Unsourced random access~(URA), proposed by Polyanskiy in~\cite{A_perspective_on}, has emerged as a novel paradigm enabling massive connectivity. 
  In such case, the base station~(BS) only needs to recover a list of transmitted messages 
  without regard for user identities.
  It allows all users to share a common codebook, thereby alleviating the burden of allocating a unique codebook for each user. 

  It is crucial to conduct information-theoretic analysis in the finite-blocklength regime to understand fundamental limits of URA. 
  Assuming the number of active users is fixed and known in advance, Polyanskiy~\cite{A_perspective_on} derived an upper bound on the 
  error probability for URA in AWGN channels.
  The fundamental limits of URA in single-receive-antenna fading channels were explored in~\cite{RAC_fading}.
  Motivated by the potential of multiple-input multiple-output~(MIMO) in enhancing spectral efficiency, non-asymptotic bounds on the minimum energy-per-bit required for reliable URA over MIMO quasi-static Rayleigh fading channels were provided in~\cite{letter}.
  The above results are established under the assumption of knowing the number of active users in advance. However, this assumption is overoptimistic because users typically have intermittent communication patterns and access the network without a grant, thereby leading to variation in the number of active users over time.
  To this end, the authors in~\cite{noKa} considered a URA scenario with a random and unknown number of active users in AWGN channels, and
  derived non-asymptotic achievability and converse bounds on misdetection and false-alarm probabilities.
  In a special random access scenario with a single potential user in binary AWGN channels, non-asymptotic bounds 
  were established in~\cite{On_joint}. 
  Moreover, some practical schemes have been proposed to approach the fundamental limits of URA~\cite{Caire2,Fasura,Duman2,SKP,ALOHA1,ALOHA2}.

  Despite various efforts have been devoted to the literature, the fundamental limits of URA with a random and unknown number of active users are still not well understood, particularly in MIMO fading channels,
  which will be explored in this work.
  First, to cope with the uncertainty of the sparsity level, we estimate the number of active users and analyze estimation performance in both non-asymptotic and asymptotic regimes.
  Then, we delve into data detection. 
  We derive non-asymptotic achievability and converse bounds on the minimum energy-per-bit required by each active user to transmit $J$~bits with blocklength~$n$ under given misdetection and false-alarm constraints.
  Our results reveal the fundamental limits of URA with random user activity and stringent latency and energy constraints, and provide valuable benchmarks to evaluate practical schemes.
  Finally, we present extensive simulation results and draw insightful conclusions from them.

  \emph{Notation:} Throughout this paper, uppercase and lowercase boldface letters denote matrices and vectors, respectively.
  We use $\left[\mathbf{x} \right]_{m}$ to denote the $m$-th element of $\mathbf{x}$ and $\left[\mathbf{X} \right]_{m,n}$ to denote the $\left( m,n \right)$-th element of $\mathbf{X}$.
  We use $\left(\cdot \right)^{T}$, $\left(\cdot \right)^{H}$, $\left|\mathbf{X}\right|$, $\left\|\mathbf{x} \right\|_{p}$, and $\left\|\mathbf{X} \right\|_{F}$ to denote transpose,
  conjugate transpose, determinant, ${\ell}_p$-norm, and Frobenius norm, respectively.
  We use $\cdot\backslash\cdot$ to denote set subtraction and $\left| \mathcal{A} \right|$ to denote the cardinality of a set $\mathcal{A}$.
  Let $x^{+} = \max\{x, 0\}$.
  For an integer $k > 0$, let $[k] = \left\{1,\ldots,k \right\}$. For integers $k_2 \geq k_1\geq 0$, let $[k_1:k_2] = \left\{k_1,\ldots,k_2\right\}$.
  We denote $h_2(p) = -p\log_2(p)-(1-p)\log_2(1-p)$ for $0 \leq  p \leq 1$. 
  We use $\gamma\left(\cdot, \cdot\right)$, $\Gamma\left(\cdot, \cdot\right)$, and $\Gamma\left(\cdot\right)$ to denote the lower incomplete gamma function, upper incomplete gamma function, and gamma function, respectively.

\section{System Model} \label{section2}

  We consider an uplink wireless network with $K$ potential single-antenna users, among which ${\rm{K}}_a$ users are active.
  We assume ${\rm{K}}_a$ is random and unknown to the receiver, but its distribution is known in advance~\cite{Yuwei_active}.
  The probability of having exactly $K_a$ active users is denoted as $P_{{\rm{K}}_a}(K_a)$. The active user set is denoted as $\mathcal{K}_a$.
  Each active user transmits $J = \log_2 M$ bits to the BS using shared $n$ channel uses.
  All users share a common codebook $\mathbf{X} = \left[\mathbf{x}_{1},\ldots, \mathbf{x}_{M}\right] \in \mathbb{C}^{n\times M}$.

  We consider MIMO quasi-static Rayleigh fading channels, where the BS is equipped with $L$ antennas.
  We assume neither the BS nor users know the instantaneous channel state information~(CSI) in advance, but they know the distribution. The received signal of the $l$-th antenna at the BS is given by
  \begin{equation} \label{eq_yl}
    \mathbf{y}_l = {\sum}_{k\in{\mathcal{K}}_a}{h}_{k,l} \mathbf{x}_{W_k} +\mathbf{z}_l \in \mathbb{C}^{n},
  \end{equation}
  where fading coefficients $\{ {h}_{k,l}: k\in\mathcal{K}_a, l\in[L]\}$ are i.i.d. $\mathcal{CN}(0,1)$ distributed;
  additive noises $\{ \mathbf{z}_l: l\in[L]\}$ are i.i.d. $\mathcal{CN}(\mathbf{0}, \mathbf{I}_n)$ distributed;
  $\mathbf{x}_{W_k}$ denotes the codeword transmitted by active user~$k$;
  and the message $W_{k}$ is chosen uniformly at random from $[M]$.
  The received signal $\mathbf{Y} = \left[ \mathbf{y}_1, \ldots, \mathbf{y}_L \right] $ is given by
  \begin{equation} \label{eq_y}
    \mathbf{Y} = \mathbf{X}\boldsymbol{\Phi}\mathbf{H}+\mathbf{Z}
    \in \mathbb{C}^{n\times L}  ,
  \end{equation}
  where $\mathbf{H} \! = \! \left[\mathbf{h}_1, \ldots, \mathbf{h}_L \right] \in \mathbb{C}^{K\times L}$ with $\mathbf{h}_l = \left[ {h}_{1,l}, \ldots,{h}_{K,l} \right]^T$;
  $\mathbf{Z} = \left[\mathbf{z}_1, \ldots, \mathbf{z}_L \right] \in \mathbb{C}^{n\times L}$;
  and $\boldsymbol{\Phi} \in \{0,1\}^{M \times K}$ contains at most one ``1'' in each column since each user transmits one message if it is active or zero messages if it is inactive.


  We provide the notion of a URA code for scenarios with random and unknown number $ {\rm K}_{ a} $ as follows \cite{noKa}:
  \begin{defi}\label{defi1}
    Let $\mathcal{X}$ and $\mathcal{Y}$ denote the input and output~alphabet, respectively. An $(n, M, \epsilon_{\mathrm{MD}}, \epsilon_{\mathrm{FA}}, P)$ URA code includes
    \begin{enumerate}
      \item
          An encoder $\emph{f}_{\text{en}}: [M] \mapsto \mathcal{X}$ that maps the message $W_k \in [M]$ to a codeword $\mathbf{x}_{W_k} \in \mathcal{X}$.
          The codewords in $\mathcal{X}$ satisfy the maximum power constraint
          \begin{equation}\label{eq:power_constraint}
            \left\|\mathbf{x}_{m}\right\|_{2}^{2} \leq nP,\;\;\;\; \forall m\in[M].
          \end{equation}
      \item
          A decoder $\emph{g}_{\text{de}}: \mathcal{Y} \mapsto \binom{[M]}{|\hat{\mathcal{W}}|}$ that satisfies the constraints on the per-user probability of misdetection and the per-user probability of false-alarm as follows:
          \begin{equation} 
            P_{\mathrm{MD}} \! = \! \mathbb{E} \! \left[ 1 \!\left[ \mathrm{K}_a \geq 1 \right]
            \cdot \frac{1}{\mathrm{K}_a}   {\sum}_{k\in {\mathcal{K}_a}} \! \mathbb{P} \! \left[ W_{k} \!\notin\! \hat{\mathcal{W}}  \right]\right] \;\!\!\leq\;\!\! \epsilon_{\mathrm{MD}},
          \end{equation}
          \begin{equation} 
            P_{\mathrm{FA}} \!=\! \mathbb{E} \bigg[ 1 \!\left[ |\hat{\mathcal{W}}| \geq 1 \right]
            \cdot  \frac{1}{ |\hat{\mathcal{W}}| } {\sum}_{ \hat{w} \in \hat{\mathcal{W}} } \mathbb{P} \!\left[ \hat{w}\;\!\! \notin \;\!\! \mathcal{W} \right] \bigg] \leq \epsilon_{\mathrm{FA}},
          \end{equation}
          where $\mathcal{W} $ denotes the set of transmitted messages and $\hat{\mathcal{W}}$ denotes the set of decoded messages of size $ | \hat{\mathcal{W}}  | \leq K$.
    \end{enumerate}
  \end{defi}

  Let $E_b = \frac{nP}{J}$ denote the energy-per-bit. The minimum required energy-per-bit is denoted as $E^{*}_{b}(n, M, \epsilon_{\mathrm{MD}}, \epsilon_{\mathrm{FA}}) = \inf \left\{E_b: \exists (n, M, \epsilon_{\mathrm{MD}}, \epsilon_{\mathrm{FA}}, P) \text{ code} \; \right\}$. 

\section{Estimation of the Number of Active Users}\label{Section:Ka}

  To cope with the uncertainty of the sparsity level, a feasible solution is to first estimate the number of active users based on $\mathbf{Y}$ and then choose the decoded list size from a set around the estimated value.
  In this part, we conduct theoretical analysis on estimation performance. 
  Theorem~\ref{Theorem_Ka} provides a non-asymptotic upper bound on the error probability 
  of estimating $\!K_a\!$ as~$\!K'_a$. 
\begin{Theorem} \label{Theorem_Ka}
  Assume $K_a$ is fixed and unknown in advance. The error probability of estimating $K_a$ as $K'_a\neq K_a$ satisfies
    \begin{equation}\label{eq_noCSI_noKa_pKa_Kahat}
      \mathbb{P} \left[ K_a  \to K'_{a} \right]
      \leq \min_{0<P'\leq P}  \left\{ p_{K_a\to K'_a, {\rm{new}} } + p_{0,K_a} \right\} ,
    \end{equation}
    \vspace{-0.4cm}
    \begin{align}
       p_{K_a\to K'_a, {\rm{new}}} =
       & \min_{ \tilde{K}_a \in \left[0 : K\right],
       \tilde{K}_a \neq K'_a }
       \big\{ 1  \big[ K'_a < \tilde{K}_a \big]
       p_{K'_{a},1}  \notag\\
       & \;\;\;\;\; \;\;\;\;\;   \;\;\;\;\;  \;\;\;\;\; \;\;
       +   1 \big[ K'_a > \tilde{K}_a \big] p_{K'_a,2} \big\} , 
    \end{align}
    \vspace{-0.3cm}
    \begin{equation}
      p_{K'_a,1} \!=\! \mathbb{E}
      \! \left[ \min_{\rho\geq0} \exp\!\left\{  \rho nL\!\left( 1\!+\!\tilde{K}'_a  P^{\prime} \right)
      \!-\!L \ln \! \left|  \mathbf{I}_n \!+\! \rho\mathbf{F} \right| \right\} \right] \!,
    \end{equation}
    \vspace{-0.3cm}
    \begin{align}
      p_{\;\!\!K'_a\;\!\!,2} \!=\! \mathbb{E}
      \bigg[\;\!\!
      \min_{\! 0\leq \rho < \;\!\!{1}\;\!\!/\;\!\!({1\;\!\!+\;\!\!\lambda'_{1}}) }
      \!\!\! \exp\!\left\{\! -\rho nL \big( 1\!+\!\tilde{K}'_a P^{\prime}  \big)
      \!-\!\;\!\!L \!\ln \!\left|  \mathbf{I}_n \!\!-\! \;\!\!\rho\mathbf{F} \right|  \!\right\} \;\!\!\;\!\!\!\bigg] \! .    
    \end{align}
    Here, $p_{0,K_a}= {K_a\Gamma \left(n, \frac{nP}{P'}\right)} \big/{\Gamma\left(n\right)}$ when the columns of $\mathbf{C} \in \mathbb{C}^{n\times M}$ are i.i.d. $\mathcal{CN} \left(0,P'\mathbf{I}_{n}\right)$ distributed,
    and $p_{0,K_a}=0$ when columns of $\mathbf{C}$ are drawn uniformly i.i.d. from a sphere of radius $\sqrt{nP'}=\sqrt{nP}$;
    $\tilde{K}'_a = \big(K'_a+\tilde{K}_a\big)\big/ \;\! 2$;
    $\mathbf{F} = \mathbf{I}_n + \mathbf{C}\boldsymbol{\Lambda}_{ \mathcal{K}_a }\mathbf{C}^H$;
    and ${\boldsymbol{\Lambda}}_{\mathcal{K}_a} \in \mathbb{N}^{M\times M}$ denotes a diagonal matrix with $\left[ {\boldsymbol{\Lambda}}_{\mathcal{K}_a} \right]_{m,m} = i$ if the $m$-th codeword is selected by $i$ users;
    and $\lambda'_1$ denotes the maximum eigenvalue of $\mathbf{C}\boldsymbol{\Lambda}_{ \mathcal{K}_a }\mathbf{C}^H$.
    \begin{IEEEproof}
    See Appendix~\ref{Appendix_proof_Ka} in the supplementary file.
    \end{IEEEproof}
\end{Theorem}

  To obtain Theorem~\ref{Theorem_Ka}, we construct a common codebook $\mathbf{C}$ with codewords i.i.d. $\mathcal{CN} \left(0,P'\mathbf{I}_{n}\right)$ distributed or drawn uniformly from a sphere of radius $\sqrt{nP'}$. The total variation distance between the measure with power constraint~\eqref{eq:power_constraint} and the new one using codebook $\mathbf{C}$ is bounded by $p_{0,K_a}$ in~\eqref{eq_noCSI_noKa_pKa_Kahat}. Then, we utilize an energy-based detector. Specifically, $K_a$ is estimated by searching an element $\tilde{K}_a\in [0:K]$ with the minimum distance $\big| \left\| \mathbf{Y} \right\|_{F}^{2} - nL \big( 1 + \tilde{K}_a P' \big) \big|$.
  The derived bound on $\mathbb{P} \left[ K_a  \to K'_{a} \right]$ decays exponentially with an increase in the number of BS antennas and eventually vanishes. However, for a fixed $L$, it reaches a plateau as power $P$ and blocklength $n$ increase.
  This is because according to the law of large numbers, adding more antennas can bring the observation $\left\| \mathbf{Y} \right\|_F^2$ closer to its mean $ nL \left( 1 + K_a P' \right)$, consequently enhancing estimation performance.
  However, increasing $P$ and $n$ cannot always work since only a small number of fading coefficients are involved 
  in quasi-static fading channels.
  From a communication perspective, increasing $P$ is ineffective in reducing multi-user interference~(MUI) and channels of some users may remain in outage even if $n\to\infty$,
  hindering a consistent decrease in error probability.
  The following Corollary~\ref{Theorem_Ka_asymP} and Corollary~\ref{Theorem_Ka_asymn} show the limit of the upper bound on $\mathbb{P} \;\!\!\left[ K_a  \! \to  \!K'_{a} \right]$ in the case of $P\;\!\!\to\;\!\!\infty$ and $n\;\!\!\to\;\!\!\infty$, respectively.

\begin{Corollary} \label{Theorem_Ka_asymP}
  When $ P\to\infty$, $\mathbb{P} \;\!\!\left[ K_a  \! \to  \!K'_{a} \right]\!$ is bounded~as
    \begin{align} 
       \mathbb{P} \left[ K_a  \to K'_{a} \right]
       \leq
       & \min_{ \tilde{K}_a \in \left[0 : K\right],
       \tilde{K}_a \neq K'_a }
       \big\{ 1  \big[ K'_a < \tilde{K}_a \big]
       p_{K'_{a},1}  \notag\\
       & \;\;\;\;\; \;\;\;\;\;   \;\;\;\;\;  \;\;\;\;\; \;\;
       +   1 \big[ K'_a > \tilde{K}_a \big] p_{K'_a,2} \big\} ,
    \end{align}
    \vspace{-0.1cm}
    \begin{equation}
      p_{K'_a,1} \!=\! \mathbb{E}
      \!  \left[ \min_{\tilde{\rho}\geq0} \exp \!\left\{ \!\!\; \tilde{\rho} nL \tilde{K}'_a
      \!-\!L \!\ln \! \left|  \mathbf{I}_n \!\!+\! \tilde{\rho}  \tilde{\mathbf{C}}\boldsymbol{\Lambda}_{ \mathcal{K}_a } \!\tilde{\mathbf{C}}^H  \right|  \right\}  \right] \! ,   
    \end{equation}
    \begin{equation}
      p_{K'_a,2} \! = \! \mathbb{E}
      \bigg[\!
      \min_{ 0\leq \tilde{\rho} < 1/{ \tilde{\lambda}'_{1} } }
      \;\!\! \!\exp \!\left\{ \!-\tilde{\rho} nL \tilde{K}'_a
      \;\!\! \!-\! L \;\!\! \ln \!\left|  \mathbf{I}_n \;\!\! \!- \!\tilde{\rho} \tilde{\mathbf{C}}\boldsymbol{\Lambda}_{ \mathcal{K}_a }\!\tilde{\mathbf{C}}^H \right|  \right\} \;\!\! \bigg]   .    
    \end{equation}
    Here, the columns of $\tilde{\mathbf{C}} \in \mathbb{C}^{n\times M}$ satisfy $\mathbb{E} [ \left\| \tilde{\mathbf{c}}_m \right\|_2^2  ]=n$ for $m \in [M]$;
    $\tilde{K}'_a$ and ${\boldsymbol{\Lambda}}_{\mathcal{K}_a}$ are given in Theorem~\ref{Theorem_Ka};
    and $\tilde{\lambda}'_1$ denotes the maximum eigenvalue of the matrix $\tilde{\mathbf{C}}\boldsymbol{\Lambda}_{ \mathcal{K}_a }\tilde{\mathbf{C}}^H$.
\end{Corollary}
\begin{Corollary} \label{Theorem_Ka_asymn}
  In the case of $n\to\infty$, the error probability of estimating $K_a$ as $K'_a\neq K_a$ can be bounded as
    \begin{align} 
       \mathbb{P} \left[ K_a \! \to \! K'_{a} \right]
       \leq\frac{\binom{K_a}{2}}{M}\! &+\!  \min_{ \tilde{K}_a \in \left[0 : K\right],
       \tilde{K}_a \neq K'_a } \!
       \big\{ 1 \!\big[ K'_a < \tilde{K}_a \big]
       p_{ K'_{a},1} \notag\\
       & \;\;\;\;\;  \;\;\;\;\;
       + 1 \big[ K'_a > \tilde{K}_a \big] p_{ K'_a,2} \big\} ,
    \end{align}
    \begin{equation}
      p_{K'_a,1}
      \!=\!
      \begin{cases}
      \exp \!\left\{ \!K_a L \!\left( 1 \!-\! \frac{\tilde{K}'_a}{K_a} + \ln\! \frac{ \tilde{K}'_a }{K_a} \right)  \right\} ,
      & \!\!\!\! \text { if }  \tilde{K}'_a  \!<\! K_a\\
      1,
      & \!\!\!\! \text { if }  \tilde{K}'_a \!\geq\! K_a
      \end{cases} \!,   
    \end{equation}
    \begin{equation}
      p_{K'_a,2} \! =\!
      \begin{cases}
      \exp \!\left\{ \!K_a L\! \left( 1 \!-\! \frac{\tilde{K}'_a}{K_a} + \ln\! \frac{ \tilde{K}'_a }{K_a} \right)  \right\} ,
      & \!\!\!\! \text { if }  \tilde{K}'_a  \!>\! K_a\\
      1,
      & \!\!\!\! \text { if }  \tilde{K}'_a \!\leq\! K_a
      \end{cases} \! .  
    \end{equation}
    Here, $\tilde{K}'_a$ is defined in Theorem~\ref{Theorem_Ka}.
    \begin{IEEEproof}
  See Appendix~\ref{Appendix_proof_converse_noCSI_Gaussian_noKa} in the supplementary file.
    \end{IEEEproof}
\end{Corollary}


\section{Data Detection}\label{Section:Data}

  In this section, we conduct information-theoretic analysis for data detection in the context of URA. 
  The non-asymptotic achievability bound is given in Section~\ref{Section:Non-Asymptotic-Results-achievability} and the converse parts are presented in Section~\ref{Section:Non-Asymptotic-Results-converse}.


\subsection{Achievability Bound} \label{Section:Non-Asymptotic-Results-achievability}

  Assume the number ${\rm K}_a$ of active users in URA scenarios is random and unknown in advance. Theorem~\ref{Theorem_achievability} presents an achievability bound on the minimum energy-per-bit required by each active user to transmit $J$ bits with blocklength $n$ under fixed misdetection and false-alarm constraints over MIMO quasi-static Rayleigh fading channels.
  \begin{Theorem} \label{Theorem_achievability}
    The minimum required energy-per-bit for the URA model described in Section~\ref{section2} can be upper-bounded as
      \begin{equation}\label{eq_noCSI_noKa_EbN0}
        E^{*}_{b}(n,M,\epsilon_{\mathrm{MD}},\epsilon_{\mathrm{FA}})
        \leq E^{*}_{b,achi}
        = \inf \frac{n {P}}{J}.
      \end{equation}
    The $\inf$ is taken over all $P > 0$ satisfying
    \vspace{-0.1cm}
      \begin{align}
        & \epsilon_{\mathrm{MD}} \geq
        \min_{0< P'< P}
        \Bigg\{  p_0 +
        \sum_{K_a = \max\{K_l,1\}}^{K_u}   \!P_{{\rm{K}}_a}(K_a)\!
        \sum_{K'_a=K_l}^{K_u}
        \sum_{t \in \mathcal{T}_{K'_a} }   \notag \\
        &  \left. \frac{t\!+\!(K_a\!-\!K'_{a,u} )^{+}\!\!}{K_a} \!\min\!\left\{ \!1,  \!{\sum}_{t' \!\in\! \bar{\mathcal{T}}_{\!K'_a\!,t}}  \!\! p_{K'_a,t,t'}, p'_{K_a\to K'_a} \!\right\}
        \! \right\} \!, \label{eq_noCSI_noKa_epsilonMD}
      \end{align}
      \vspace{-0.2cm}
      \begin{align} \label{eq_noCSI_noKa_epsilonFA}
        \epsilon_{\mathrm{FA}} \!\geq \!
        &  \min_{0< P'< P}
        \!\Bigg\{  p_0 + \!\!
        \sum_{K_a=K_l}^{K_u} \!\! P_{{\rm{K}}_a}(K_a)\!\!
        \sum_{K'_a=K_l}^{K_u} \sum_{t \in \mathcal{T}_{K'_a} } \sum_{t' \in \mathcal{T}_{K'_a,t} }
          \notag\\
         & \;\;\;  \left. \frac{t' \!+\!  ( K'_{a,l}\!-\!K_a\! )^{+}}{ \hat{K}_a  }
        \min\!\left\{ 1, p_{K'_a,t,t'}, p'_{K_a \to K'_a} \right\} \!\right\}\!,
      \end{align}
      \vspace{-0.1cm}
    \begin{equation} \label{eq_noCSI_p0}
      p_0 \!= \!\!\sum_{K_a\!=0}^{K} \!P_{{\rm{K}}_a}\!(K_a) \!\left( \!\frac{\binom{K_a}{2}}{M} \!+ p_{0,K_a} \!\!\right)
      \!+ \! \!\!\sum_{K_a \notin [K_l:K_u]}   \!\!P_{{\rm{K}}_a}\!(K_a) ,
    \end{equation}
      \begin{equation}\label{eq_noCSI_noKa_estimate_Tset1}
        \mathcal{T}_{K'_a } \!=\!  \left[ 0  \!:
        \min\!\left\{K_a,K'_{a,u},M\!-\!K'_{a,l}\!-\!(K_a\!-\!K'_{a,u})^{+}\right\} \right]\! ,
      \end{equation}
      \begin{equation}\label{eq_noCSI_noKa_estimate_Tset2bar}
        \bar{\mathcal{T}}_{K'_a,t} \!=\!   \!
        \left[ \left( (K_a \!-\! K'_{a,u} )^{+} \!\!-\! (K_a\!-\!K'_{a,l} )^{+} \!+ t \right)^{+}
        \!:\! T_{upper} \right]\! ,
      \end{equation}
      \begin{align}
        T_{upper} = \min & \left\{ (K'_{a,u} \!\!-\! K_a)^{+} \!\!-\! (K'_{a,l} \!-\! K_a)^{+} \!+\! t , \right. \notag\\
        & \;\;\left.
        M-\max \left\{ K_a , K'_{a,l} \right\} \right\} ,
      \end{align}
      \begin{equation}\label{eq_noCSI_noKa_estimate_Kahat}
        \hat{K}_a = K_a - t - (K_a-K'_{a,u} )^{+} + t' + ( K'_{a,l}  - K_a )^{+} ,
      \end{equation}
    \begin{align} 
      & p_{K'_a,t,t'} \!=\!  \min_{ 0 \leq \omega \leq 1, 0 \leq \nu }\!
      \left\{ 1 \!\left[ t+(K_a\!-\!K'_{a,u})^{+} > 0 \right]
      q_{2,K'_a,t }\left(\omega,\nu\right)  \right. \notag \\
      &   \left. + 1 \!\left[ t\!+\!(K_a\!-\!K'_{a,u})^{+} \!= 0 \right] q_{2,K'_a,t,0 }\!\left(\omega,\nu\right)
      + q_{1,K'_a,t,t'}\!\left(\omega,\nu\right) \right\} \! ,
    \end{align}
    \vspace{-0.3cm}
    \begin{align}
      & q_{1,K'_a\!,t,t'} \!\left(\omega,\nu\right) \!=\! C_{\!K'_a \!,t ,t'}
       \mathbb{E} \! \left[
      \min_{ {u\geq 0,r\geq 0, \lambda_{\min} (\mathbf{B}) > 0}}\!\!
      \exp \! \left\{  L rn\nu  \!+\! b_{u,r} \right.
      \right. \notag \\
      &  \left. +\!  L \! \left( u \ln \!\left|\mathbf{F}''\right|
      \!-\! r \ln \!\left|\mathbf{F}\right|
      \!-\! u \ln \!\left| {\mathbf{F}'} \right|
      \!+\! r\omega \ln \!\left| \mathbf{F}_{1} \right|
      \!-\! \ln \!\left| \mathbf{B} \right| \right)
      \right\} \!\Big] ,   
    \end{align}
      \begin{equation}\label{eq_noCSI_noKa_estimation_B}
        \mathbf{B} = (1+r) \mathbf{I}_n - u \left( \mathbf{F}'' \right)^{-1} \mathbf{F}
        + u \left( \mathbf{F}' \right)^{-1} \mathbf{F}
        - r\omega \mathbf{F}_{ 1}^{-1} \mathbf{F} ,
      \end{equation}
    \begin{equation}\label{eq_noCSI_noKa_estimation_bur}
      b_{u,r} = -u b'' + r b + u b' -r\omega b_1,
    \end{equation}
    \begin{align}
      & q_{2,K'_a,t}\!\left(\omega,\nu\right) =\! \binom{K_a}{t\!+\!(K_a\!-\!K'_{a,u})^{+}\!}
      \min_{ \delta\geq0 }\!
      \Bigg\{ \frac{\Gamma\!\left( nL, nL \left( 1\!+\delta \right)\right)}{\Gamma\left( nL \right)} \notag\\
      &  \left. \!+\;\!   \mathbb{E} \!  \left[ \! \frac{\gamma\!\left(\! Lm, \!\frac{ nL(1+\delta)(1-\omega) - \omega\left( L\!\ln\left|\mathbf{F}_1\!\right| - b_1 \right) + L\! \ln\left|\mathbf{F}\right| - b - nL\nu }
      {\omega \prod_{i=1}^{m} \! \lambda_i^{1/m} } \! \right)}{\Gamma\left( Lm \right)} \!\right]
      \!\right\} \! ,   
    \end{align}
    \vspace{-0.1cm}
    \begin{equation}\label{eq_noCSI_noKa_estimation_q2t0}
      q_{2,K'_a,t,0} = \mathbb{E} \!\left[  \frac{\Gamma\!\left( nL, {nL\nu}/({1-\omega}) - L \!\ln \left| \mathbf{F} \right| + b \right)}{\Gamma\left( nL \right)}  \right] .
    \end{equation}
  Here, 
  $p'_{K_a\to K'_a}$ is obtained from $p_{K_a\to K'_a, {\rm{new}}}$ in Theorem~\ref{Theorem_Ka} by changing $\left[0 : K\right]$ to $\left[K_l : K_u\right]$ with $0\leq K_l \leq K_u \leq K$ and assuming there is no collision;
  $K'_{a,l}  = \max\left\{ K_l , K'_{a} -r' \right\}$ and $K'_{a,u}  = \min\left\{ K_u , K'_{a} +r' \right\}$;
  $r'$ denotes a nonnegative integer referred to as the decoding radius;
  $b=f_b\left(K_a\right)$, $b'=f_b\big({\hat{K}}_a\big)$, $b''=f_b\big( K_a \! - \! (K_a\!-\!K'_{a,u} )^{+} + ( K'_{a,l}  \!-\! K_a )^{+}  \big)$, and $b_1=f_b\left( K_a \!-\! t \!-\! (K_a\!-\!K'_{a,u})^{+} \right)$
  with $f_b(\cdot) = \ln \left( P_{ {\rm{K}}_a} \left(\cdot\right) \right) -
      \ln  \binom{M}{ \cdot }$;
  $\mathbf{F} = \mathbf{I}_n + \mathbf{C}\boldsymbol{\Gamma}_{ \mathcal{W} }\mathbf{C}^H$,
    $\mathbf{F}_1  = \mathbf{I}_n + \mathbf{C}\boldsymbol{\Gamma}_{ \mathcal{W} \backslash \mathcal{W}_1 }\mathbf{C}^H$,
    $\mathbf{F}'  = \mathbf{I}_n + \mathbf{C}\boldsymbol{\Gamma}_{ \mathcal{W} \backslash \mathcal{W}_1 \cup \mathcal{W}_2 }\mathbf{C}^H$,
    and $\mathbf{F}''  = \mathbf{I}_n + \mathbf{C}\boldsymbol{\Gamma}_{ \mathcal{W} \backslash \mathcal{W}_{1,1} \cup \mathcal{W}_{2,1} }\mathbf{C}^H$;
  $\mathcal{W}$ is an arbitrary $K_a$-subset of $[M]$;
  $\mathcal{W}_{1}$ is an arbitrary $t+(K_a-K'_{a,u})^{+}$-subset of $\mathcal{W}$, which is divided into two subsets $\mathcal{W}_{1,1}$ and $\mathcal{W}_{1,2}$ of size $(K_a-K'_{a,u})^{+}$ and $t$, respectively;
  $\mathcal{W}_{2}$ is an arbitrary $t'+(K'_{a,l}-K_a)^{+}$-subset of $[M] \backslash \mathcal{W}$; $\mathcal{W}_{2,1}$ is an arbitrary subset of $\mathcal{W}_2$ of size $(K'_{a,l}-K_a)^{+}$;
  for any subset $S \subset [M]$, the matrix ${\boldsymbol{\Gamma}}_{S} = \operatorname{diag} \left\{ {\boldsymbol{\gamma}}_{S} \right\}\in \left\{ 0,1 \right\}^{M\times M}$, where $\left[ {\boldsymbol{\gamma}}_{S} \right]_{i} = 1$ if $i \in S$ and $\left[ {\boldsymbol{\gamma}}_{S} \right]_{i} = 0$ otherwise;
  $\lambda_1,\ldots, \lambda_m$ denote the non-zero eigenvalues of $\mathbf{F}_1^{-1} \mathbf{C}\boldsymbol{\Gamma}_{\mathcal{W}_1}\mathbf{C}^H$ with $m=\min\left\{ n, t+(K_a-K'_{a,u})^{+} \right\}$;
  and $C_{K'_a,t,t'} = \binom{K_a}{ t+(K_a\!-\!K'_{a,u})^{+} } \binom{M-K_a}{t'+(K'_{a,l}\!-\!K_a)^{+}} $.
  \begin{IEEEproof}
    See Appendix~\ref{Appendix_proof_achievability} in the supplementary file.
    \end{IEEEproof}
  \end{Theorem}

  When the number of active users is known in advance, the BS typically selects it as the decoded list size~\cite{A_perspective_on}.
  In scenarios with random and unknown ${\rm K}_a$, selecting an appropriate list size becomes a challenge.
  In this work, we apply the maximum \emph{a posteriori}~(MAP) criterion to decode with the list size in a set around the estimated number of active users.
  Compared to the maximum likelihood criterion used in~\cite{noKa}, the MAP criterion tightens the bound by involving prior distribution of~${\rm K}_a$.

  When each user has an individual codebook as considered in~\cite{TIT}, the BS aims to recover transmitted codewords from $K$ codebooks and at most one codeword can be returned from each codebook, which can be formulated as a block sparse support recovery problem with the byproduct of user identity recovery.
  In contrast, data detection is a sparse support recovery problem without block constraints in URA scenarios.
  Moreover, compared with \cite{TIT}, where the BS needs to check whether the transmitted message of each user matches its decoded counterpart, misdetection and false-alarm probabilities are more relaxed in URA scenarios since they are determined by set differences between transmitted messages and decoded messages.
  Thus, compared to scenarios with individual codebooks, the URA paradigm does not recover user identities, but achieves higher spectral and energy efficiency due to reduced search space for decoding and relaxed error probabilities.


  Theorem~\ref{Theorem_achievability} can be adapted to both cases of using codewords with symbols i.i.d. $\mathcal{CN}(0,P')$ distributed and using codewords uniformly distributed on a sphere of radius $\sqrt{nP'}$.
  For the Gaussian random coding scheme, $P'$ must be smaller than $P$ to guarantee a low total variation distance $p_{0,K_a}$.
  However, when using codewords distributed on a sphere, $P'$ can be equal to $P$,
  providing an advantage in the non-asymptotic~regime.

\subsection{Converse Bound} \label{Section:Non-Asymptotic-Results-converse}

  We provide three converse bounds on the minimum required energy-per-bit for URA in MIMO quasi-static Rayleigh fading channels with random and unknown ${\rm K}_a$.
  Theorem~\ref{Theorem_converse_single} presents a single-user type converse bound. It is obtained by casting a URA code as a single-user code with list decoding via assuming the activities of all users and transmitted messages and CSI of ${\rm K}_a-1$ active users are known \emph{a priori}.
  \begin{Theorem} \label{Theorem_converse_single}
   The minimum required energy-per-bit for the URA model described in Section~\ref{section2} can be lower-bounded as
    \begin{equation} \label{eq:P_tot_conv_singleUE_EbN0}
       E^{*}_{b}(n, M, \epsilon_{\mathrm{MD}}, \epsilon_{\mathrm{FA}})
      \geq \inf \frac{nP}{J}.
    \end{equation}
    Here, the $\inf$ is taken over all $P > 0$ satisfying that
      \begin{equation} \label{eq:P_tot_conv_singleUE1}
        J  - \log_2  \! \hat{K}_a  \! \leq \!
         - \log_2  \! \mathbb{P}\! \left[ \chi^2(2L) \geq (1+(n + 1)P)    r_{K_a}
        \right] \!  ,
      \end{equation}
      where $\hat{K}_a\in [K]$ denotes the size of the decoded list when there are $K_a$ active users and $r_{K_a}$ is the solution of
      \begin{equation}\label{eq:P_tot_conv_singleUE2}
        \mathbb{P} \left[ \chi^2(2L) \leq r_{K_a} \right] = \epsilon_{K_a} ,
      \end{equation}
      under the constraints that for any subset $S \subset \mathcal{K}$,
      \begin{equation}\label{eq:P_tot_conv_singleUE3}
        {\sum}_{K_a \in S}  P_{{\rm K}_a}(K_a) \; \epsilon_{K_a} \leq \epsilon_{\mathrm{MD}},
      \end{equation}
      \begin{equation} \label{P_tot_conv_noCSI_FA}
      \sum_{K_a \in S}  P_{{\rm K}_a}(K_a) \max \bigg\{ \frac{\hat{K}_a-K_a}{\hat{K}_a} , 0 \bigg\} \leq \epsilon_{\mathrm{FA}} .
    \end{equation}
  \end{Theorem}

  Theorem~\ref{Theorem_converse_single_noKa} given below is also a single-user type converse bound, where we assume the activities of $K-1$ users, as well as transmitted codewords and CSI of active users among them, are known in advance.
  Different from Theorem~\ref{Theorem_converse_single},
  Theorem~\ref{Theorem_converse_single_noKa} takes the uncertainty of user activity into account, 
  but only applicable when ${\rm K}_a$ follows the Binomial distribution.
  \begin{Theorem} \label{Theorem_converse_single_noKa}
    Assume that ${\rm K}_a\sim {\rm{Binom}}(K,p_a)$. 
    The minimum required energy-per-bit for the URA model described in Section~\ref{section2}
    can be lower-bounded as
    \begin{equation} \label{eq:P_tot_conv_singleUE_EbN0_noKa}
      E^{*}_{b}(n,M,\epsilon_{\rm MD},\epsilon_{\rm FA}) \geq \inf \frac{nP}{J}.
    \end{equation}
    Here, the $\inf$ is taken over all $P > 0$ satisfying that
      \begin{equation} \label{eqR:beta_conv_AWGN_singleUE_1e}
        \frac{M}{K} \leq 
        \frac{ \min\left\{ 1,  { \epsilon_{\rm{FA}} }/{(1-p_a)} \right\} }{ \mathbb{P}\left[ \chi^2(2L) \geq (1+(n+1)P)r
        \right] } ,
      \end{equation}
    where $r$ is the solution of $\mathbb{P} \left[ \chi^2(2L) \leq r \right] \! = \! \min\left\{ 1,  { \epsilon_{\rm{MD}} } / {p_a} \right\}$.
    \begin{IEEEproof}
      See Appendix~\ref{Appendix_proof_converse_single_noKa} in the supplementary file.
    \end{IEEEproof}
  \end{Theorem}



  Theorems~\ref{Theorem_converse_single} and \ref{Theorem_converse_single_noKa} are applicable to all codes.
  However, they can be loose in large ${\rm K}_a$ regime since they are derived based on knowledge of transmitted messages and CSI of ${\rm K}_a-1$ active users.
  This knowledge is difficult to obtain because MUI is severe in this case.
  To this end, we consider the case without this knowledge in Theorem~\ref{Theorem_converse_noCSI_Gaussian_cKa}, and apply the Fano inequality to directly address the challenge of massive access.
  Since it is tricky to analyze, we make a stronger assumption that the codebook has i.i.d. entries with mean $0$ and variance $P$. 
  \begin{Theorem}\label{Theorem_converse_noCSI_Gaussian_cKa} 
    Assuming $\epsilon_{\mathrm{MD}} \leq \frac{M}{1+M}$ and the codebook has i.i.d. entries with mean $0$ and variance $P$,
    the minimum required energy-per-bit for the URA model in Section~\ref{section2}~satisfies
    \begin{equation} \label{eq:P_tot_conv_EbN0_Gaussian}
      E^{*}_{b}(n, M, \epsilon_{\mathrm{MD}}, \epsilon_{\mathrm{FA}} )  \geq \inf \frac{nP}{J}.
    \end{equation}
    Here, the $\inf$ is taken over all $P > 0$ satisfying \eqref{P_tot_conv_noCSI_FA} and
    \begin{align}
    & \left( \mathbb{P}[ {\rm K}_a \! \in \!S] - \epsilon_{\rm MD} \right) J
    - h_2(\epsilon_{\rm MD})
    \!-\! {\sum}_{K_a \in S} P_{{\rm K}_a}\!(K_a) \log_2 \!\hat{K}_a  \notag\\
    & \leq {\sum}_{K_a \in S} P_{{\rm K}_a}(K_a) \left(
    \frac{ nL  \log_2 ( 1 + K_aP )}{K_a}  \right. \notag\\
    & \;\;\;\;\; \;
    - \frac{ L  \big( 1 - {\binom{K_a}{2}}/{M} \big) \mathbb{E}  \left[  \log_2 \left| \mathbf{I}_{n}  +  {\mathbf{X}}_{K_a}   {\mathbf{X}}_{K_a}^{ H}  \right|  \right]  }{K_a} \Bigg)  , \label{P_tot_conv_noCSI}
    \end{align}
    for any subset $S \subset \mathcal{K}$, where 
    ${\mathbf{X}}_{K_a} \in \mathbb{C}^{n\times K_{a}}$ has i.i.d. entries with mean $0$ and variance $P$.
  \begin{IEEEproof}
  See Appendix~\ref{Appendix_proof_converse_noCSI_Gaussian_noKa} in the supplementary file.
  \end{IEEEproof}
  \end{Theorem}

  A converse bound for URA in AWGN channels with a random and unknown number of active users was provided in~\cite{noKa}. However, it is an ensemble converse bound and specific to a two-step decoding scheme.
  In this work, we remove both the two restrictions in Theorems~\ref{Theorem_converse_single} and \ref{Theorem_converse_single_noKa}, and remove the second restriction in Theorem~\ref{Theorem_converse_noCSI_Gaussian_cKa}.
  The derived results in Theorems~\ref{Theorem_converse_single}, \ref{Theorem_converse_single_noKa}, and \ref{Theorem_converse_noCSI_Gaussian_cKa} complement each other and collectively characterize the lower bound on the minimum required energy-per-bit in the considered URA channels.



\section{Numerical Results} \label{Section:simulation}

  In this section, we provide numerical results to evaluate the derived bounds.
  The expectations in non-asymptotic results are evaluated using the Monte Carlo method with $2000$ samples.

\begin{figure}
	\centering
    \subfigure[]{\includegraphics[width=0.48\linewidth]{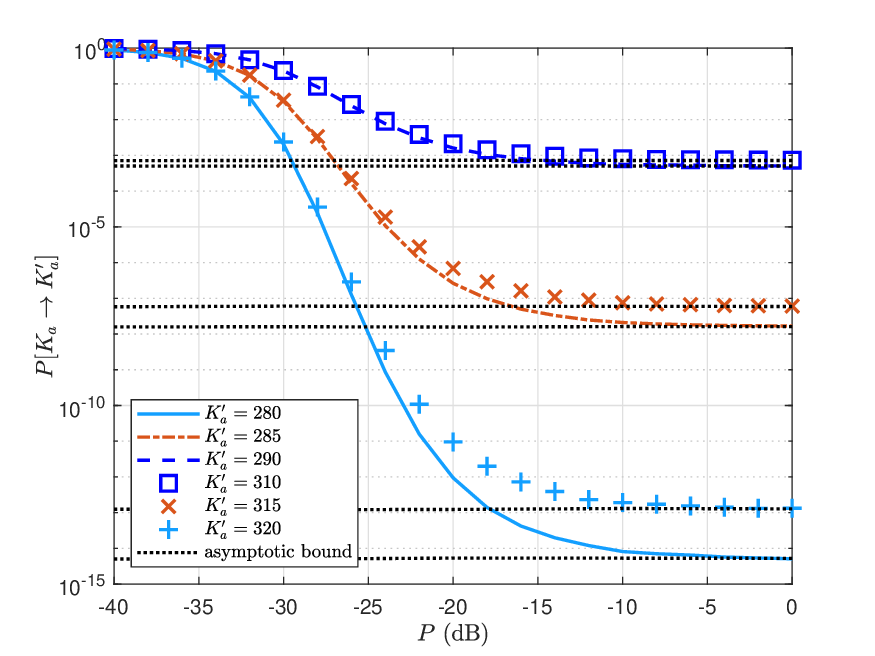}
		\label{fig:Ka_estimateP}}
    \subfigure[]{\includegraphics[width=0.48\linewidth]{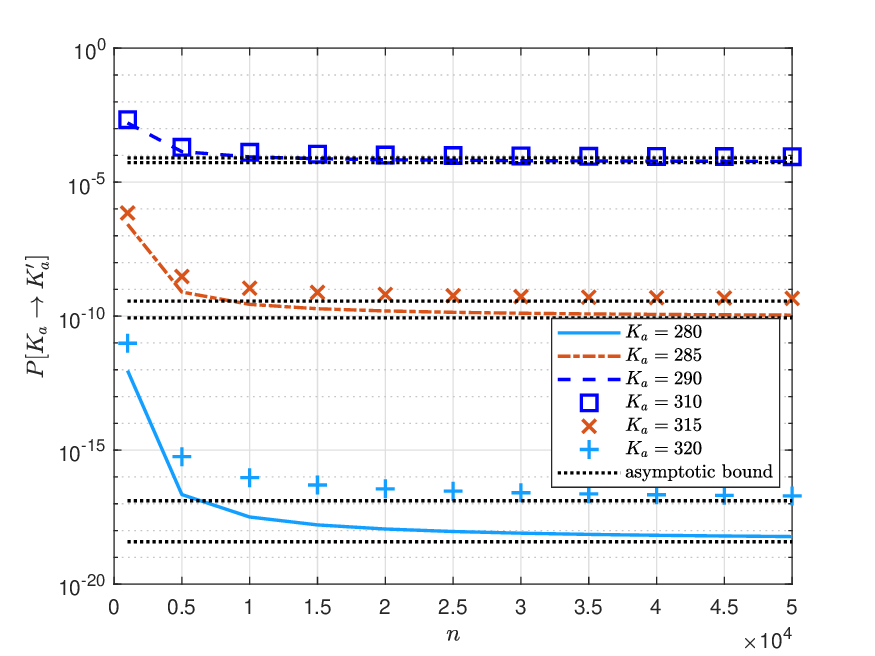}
		\label{fig:Ka_estimaten}}
	\caption{The achievability bound on $\mathbb{P}\left[ K_a \to K'_a \right]$ with $K_a=300$ and $K=600$:
(a)~$\mathbb{P}\left[ K_a \to K'_a \right]$ versus $P$ for different values of $K'_a$ with $n=1000$ and~$L=64$;
(b)~$\mathbb{P}\left[ K_a \to K'_a \right]$ versus $n$ 
with $L=64$ and $P=-20$~dB.
}\label{fig:Ka_estimate_all}
\end{figure}

  In Fig.~\ref{fig:Ka_estimate_all}, we assume there are $K_a=300$ active users and present the non-asymptotic upper bound~(in Theorem~\ref{Theorem_Ka}) on the error probability $\mathbb{P}\left[ K_a \to K'_a \right]$ of estimating $K_a$ as $K'_a\neq K_a$.
  It is shown that as power $P$ and blocklength $n$ increase, the estimation error probability initially decreases but eventually converges to the error floor provided in Corollary~\ref{Theorem_Ka_asymP} and Corollary~\ref{Theorem_Ka_asymn}, as predicted in Section~\ref{Section:Ka}.

  \begin{figure}
    \centering
    \includegraphics[width=0.78\linewidth]{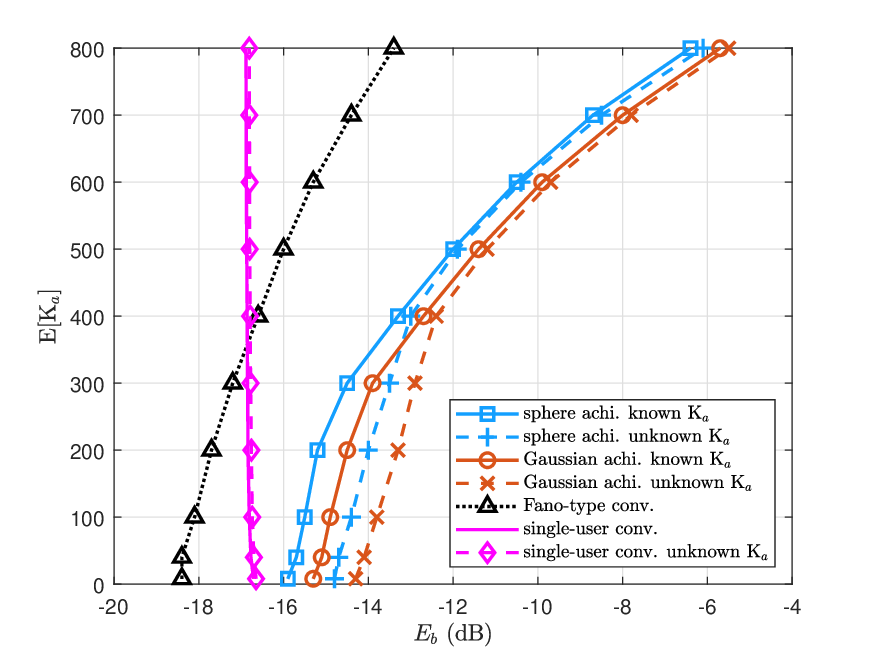}\\
    \caption{The mean of ${\rm K}_a$ versus energy-per-bit with ${\rm K}_a \sim {\rm Binom}(K,0.5)$, $n=1000$, $J = 100$~bits, $L=128$, and $\epsilon_{\rm MD} = \epsilon_{\rm FA} = 0.001$.}
  \label{fig:randomKa_EbKa}
  \end{figure}

  In Fig.~\ref{fig:randomKa_EbKa},
  we compare the derived bounds on the minimum required energy-per-bit for different values of $\mathbb{E}[{\rm K}_a]$.
  The achievability bound for the case with random and unknown ${\rm K}_a$ corresponds to Theorem~\ref{Theorem_achievability}.
  When ${\rm K}_a$ is random and known, the achievability bound is obtained from Theorem~\ref{Theorem_achievability} by selecting ${\rm K}_a$ as the decoded list size.
  The single-user converse bound and Fano-type converse bound (using Gaussian codebook) correspond to Theorem~\ref{Theorem_converse_single} and Theorem~\ref{Theorem_converse_noCSI_Gaussian_cKa}, respectively, which are converse no matter ${\rm K}_a$ is known or not. 
  The single-user converse bound with 
  unknown ${\rm K}_a$ corresponds to Theorem~\ref{Theorem_converse_single_noKa}.
  Numerical results confirm the tightness of our bounds, with the gap between achievability and converse bounds less than $4$~dB when $\mathbb{E}[{\rm K}_a]\leq 400$. 
  As explained in Section~\ref{Section:Non-Asymptotic-Results-converse}, the single-user bound dominates in small $\mathbb{E}[{\rm K}_a]$ regime, while the Fano-type bound dominates when $\mathbb{E}[{\rm K}_a]$ is large.
  Using codewords distributed on a sphere outperforms the Gaussian random coding scheme by about $0.6$~dB. 
  On the achievability side, the extra required energy-per-bit due to the uncertainty of the sparsity level is about $1.1$~dB when $\mathbb{E}[{\rm K}_a] \leq 300$. However, in large $\mathbb{E}[{\rm K}_a]$ regime, it reduces to be less than $0.3$~dB since MUI, instead of the lack of knowledge of ${\rm K}_a$, becomes the bottleneck.
  The gap between single-user converse bounds with and without known ${\rm K}_a$ is slight since this gap only reflects the energy efficiency penalty for a single~user.

  \begin{figure}
    \centering
    \includegraphics[width=0.78\linewidth]{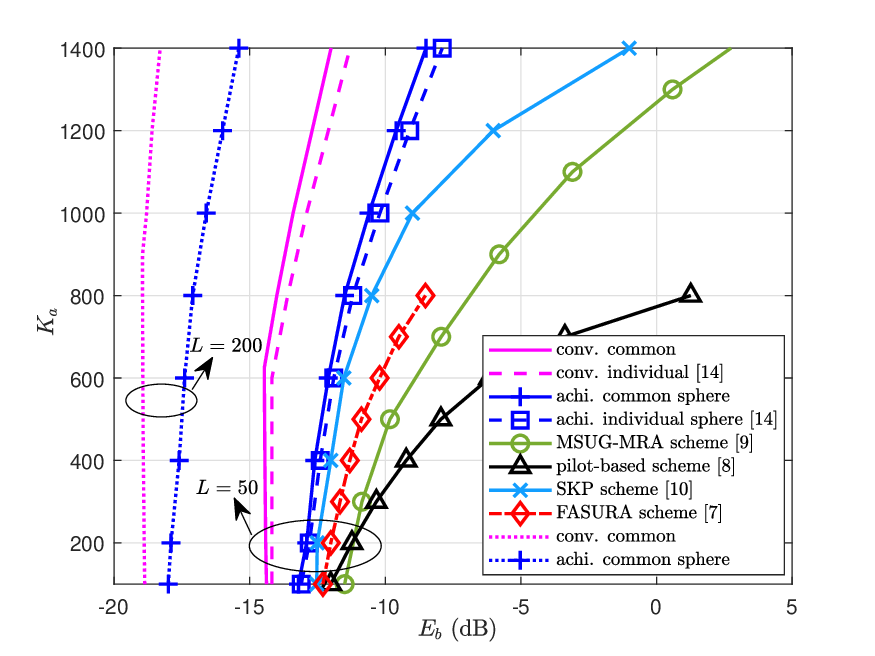}\\
    \caption{Comparison of existing schemes and theoretical bounds in the case of $n=3200$, $J = 100$~bits, $L\in\{50,200\}$, and $\epsilon_{\rm MD} = \epsilon_{\rm FA} = 0.025$. (Let $K=K_a/0.4$ for the scenario with individual codebooks.)}
  \label{fig:EbL50}
  \end{figure}
  In Fig.~\ref{fig:EbL50}, we compare theoretical bounds in scenarios with individual codebooks and a common codebook, as well as practical schemes proposed in~\cite{Caire2,Fasura,Duman2,SKP}, in the case with a fixed and known number $K_a$ of active users.
  The bounds for scenarios with individual codebooks are plotted based on Corollary~7 and Theorem~9 in~\cite{TIT} by changing Gaussian codebook in the achievability bound to the one with each codeword distributed on a sphere.
  We observe that increasing the number of BS antennas substantially decreases the minimum energy-per-bit required to support $K_a$ active users, and increases the threshold of the number of active users below which the minimum required energy-per-bit is almost a constant.
  Compared to scenarios with individual codebooks, the URA paradigm achieves higher spectral and energy efficiency, 
  with the saved energy-per-bit becoming more obvious as $K_a$ increases.
  It is shown that the SKP scheme proposed in~\cite{SKP} outperforms other schemes, but still exhibits a large gap to our bound when $K_a$ is large,
  calling for more advanced schemes that perform energy-efficiently in this case.

\section{Conclusion} \label{Section:conclusion}




In this paper, we shed light on the fundamental limits of URA with a random and unknown number of active users in MIMO quasi-static Rayleigh fading channels.
To cope with the uncertainty of the sparsity level, we estimated the number of active users and derived an upper bound on the error probability of estimating $K_a$ as $K'_a$. The error floor in the case of $P\to\infty$ and $n\to\infty$ was provided, respectively.
Then, we derived non-asymptotic achievability and converse bounds on the minimum energy-per-bit required by each active user to reliably transmit $J$ bits. 
Simulation results reveal the superiority of using codewords distributed on a sphere over a Gaussian codebook in the non-asymptotic regime, as well as the advantage of URA over random access with individual codebooks. 
Moreover, as $\mathbb{E}\left[ {\rm K}_a \right]$ decreases, the energy efficiency penalty suffering from the lack of 
the sparsity level becomes more obvious.
Our bounds provide benchmarks to evaluate existing schemes, which are shown to exhibit a large gap to our bound when the number of active users is large.
\IEEEtriggeratref{7}



\clearpage
The supplementary material contains the proofs of Theorem~\ref{Theorem_Ka}, Corollary~\ref{Theorem_Ka_asymn}, Theorem~\ref{Theorem_achievability}, Theorem~\ref{Theorem_converse_single_noKa}, and Theorem~\ref{Theorem_converse_noCSI_Gaussian_cKa}. The equation numbers in this supplementary material are continued from the paper.

\appendices

\section{Proof of Theorem~\ref{Theorem_Ka}} \label{Appendix_proof_Ka}

  Assume there are $K_a$ active users, which is fixed and unknown in advance. In this appendix, we derive a non-asymptotic upper bound on the error probability $\mathbb{P} \left[ K_a  \to K'_{a} \right]$ of estimating $K_a$ as $K'_a\neq K_a$.

      We utilize an energy-based detector to estimate the number $K_a$ of active users, which relies on the fact that following from the law of large numbers, $\left\|\mathbf{Y}\right\|_F^2$ concentrates around its mean when $L$ is large.
      Specifically, $K_a$ is estimated as
      \begin{equation}\label{eq:proof_noCSI_noKa_Kaestimate}
        K'_{a} = \arg \min_{\tilde{K}_a \in\left[0: K\right] }\; m(\mathbf{Y}, \tilde{K}_a),
      \end{equation}
      where the energy-based metric $m(\mathbf{Y}, \tilde{K}_a)$ is given by
      \begin{equation}\label{eq:proof_noCSI_noKa_Kaestimate_m}
        m(\mathbf{Y}, \tilde{K}_a) =\left| \left\| \mathbf{Y} \right\|_{F}^{2} - nL \left( 1 + \tilde{K}_a P^{\prime} \right) \right|,
      \end{equation}
      under the assumption that the columns of $\mathbf{C} = \left[ \mathbf{c}_1, \ldots, \mathbf{c}_M \right] \in \mathbb{C}^{n\times M}$ satisfy $\mathbb{E} [ \left\| \mathbf{c}_m \right\|_2^2  ] = nP'$ for $m \in [M]$.
  \begin{equation} \label{eq:proof_noCSI_noKa_Ka_Kahat_p0}
      \mathbb{P} \left[ K_a  \to K'_{a} \right]
      \leq \min_{0<P'\leq P}    \left\{ \mathbb{P} \left[ K_a  \to K'_{a} \right]_{\text {new}} + p_{0,K_a} \right\} .
    \end{equation}
  The term $p_{0,K_a}$ upper-bounds the total variation distance between the measures with and without power constraint as given in Theorem~\ref{Theorem_Ka}.
  Next, we aim to derive an upper bound on $\mathbb{P}  \left[ K_a   \to  K'_{a} \right]_{\text {new}} $ in the case of $K'_a < \tilde{K}_a$ and $K'_a > \tilde{K}_a$, respectively.
  Let $C_{K'_a,\tilde{K}_a}   =    (K'_a + \tilde{K}_a)/2$ and $C_{K'_a,\tilde{K}_a,P'}  =  1 + C_{K'_a,\tilde{K}_a} P^{\prime}$. Applying the energy-based metric $m(\mathbf{Y}, \tilde{K}_a)$ given in~\ref{eq:proof_noCSI_noKa_Kaestimate_m}, we have
  \begin{align}
    & \mathbb{P} \left[ K_a \to K'_{a} \right]_{\text {new}} \notag\\
    & \leq \min_{ \tilde{K}_a \in \left[0:K\right] \backslash K'_a }
    \mathbb{P}\left[m\left( \mathbf{Y}, K'_a \right)
    \leq m(\mathbf{Y}, \tilde{K}_a )\right] \\
    & =   \min_{ \tilde{K}_a \in \left[0:K\right] \backslash K'_a }
       \left\{  1  \left[ K'_a   <  \tilde{K}_a \right]
    \mathbb{P}  \left[ \left\| \mathbf{Y} \right\|_{F}^{2} \leq  nL C_{K'_a,\tilde{K}_a,P'} \right] \right\} \notag \\
    & \;\;\;\;\;\;\;\;   \left.
    +  1  \left[ K'_a  >  \tilde{K}_a \right]
    \mathbb{P}  \left[ \left\| \mathbf{Y} \right\|_{F}^{2} \geq nL C_{K'_a,\tilde{K}_a,P'} \right] \right\} . \label{eq:proof_noCSI_noKa_Ka_Kahat}
  \end{align}

  The probability $\mathbb{P} \left[ \left\| \mathbf{Y} \right\|_{F}^{2} \leq nL C_{K'_a,\tilde{K}_a,P'} \right] $ in~\eqref{eq:proof_noCSI_noKa_Ka_Kahat} can be bounded as follows:
  \begin{align}
    & \mathbb{P} \left[ \left\| \mathbf{Y} \right\|_{F}^{2} \leq nL C_{K'_a,\tilde{K}_a,P'} \right] \notag\\
    & \leq  \mathbb{E}   \left[ \min_{\rho\geq0}
    \exp \left\{ \rho nLC_{ K'_a ,\tilde{K}_a ,P'} \right\}
     \mathbb{E}
      \left[ \left.  \exp  \left\{  - \rho  \left\|\mathbf{Y}\right\|_F^{2} \right\} \right|  \mathbf{C} \right] \right] \label{eq:proof_noCSI_noKa_Ka_Kahat_Chernoff_11} \\
    & =  \mathbb{E}  \left[ \min_{\rho\geq0} \exp \left\{ \rho nLC_{K'_a,\tilde{K}_a,P'}
     -  L   \ln \left| \mathbf{I}_n    +   \rho\mathbf{F} \right| \right\} \right] \label{eq:proof_noCSI_noKa_Ka_Kahat11} ,
  \end{align}
  where $\mathbf{F}$ is given in Theorem~\ref{Theorem_Ka}; \eqref{eq:proof_noCSI_noKa_Ka_Kahat_Chernoff_11} follows from the Chernoff bound $\mathbb{P}\left[W>0\right] \leq \mathbb{E}\left[ \exp\left\{sW\right\} \right]$ for $s\geq 0$~\cite{elements_IT};
  \eqref{eq:proof_noCSI_noKa_Ka_Kahat11} follows by applying Lemma~\ref{expectation_bound} shown below to the conditional expectation in~\eqref{eq:proof_noCSI_noKa_Ka_Kahat_Chernoff_11}.
  Denote the RHS of \eqref{eq:proof_noCSI_noKa_Ka_Kahat11} as $p_{K'_a,1}$.
  \begin{Lemma}[\cite{quadratic_form1}]\label{expectation_bound}
    Assume that ${\mathbf{x}} \in \mathbb{C}^{p\times 1}$ is distributed as $ {\mathbf{x}} \sim \mathcal{CN}\left( \mathbf{0},  {\boldsymbol{\Sigma}} \right)$.
    Let $\mathbf{B}\in\mathbb{C}^{p\times p}$ be a Hermitian matrix.
    For any $\gamma$, if the eigenvalues of $\mathbf{I}_{p} - \gamma {\boldsymbol{\Sigma}}\mathbf{B}$ are positive, we have
    \begin{equation}\label{eq:lemma_expectation}
      \mathbb{E}\left[ \exp \left\{ \gamma \mathbf{x}^{H}\mathbf{B}\mathbf{x} \right\} \right] = \left|\mathbf{I}_{p} - \gamma \boldsymbol{\Sigma}\mathbf{B}\right|^{-1} .
    \end{equation}
  \end{Lemma}

  Likewise, we can derive the upper bound $p_{K'_a,2}$ on the probability $\mathbb{P}\!\left[ \left\|\mathbf{Y}\right\|_F^{2} \!\geq nLC_{K'_a,\tilde{K}_a,P'} \right]$ as given in Theorem~\ref{Theorem_Ka}.

\section{Proof of Corollary~\ref{Theorem_Ka_asymn}} \label{Appendix_proof_Ka_asymn}
    Assume all users share a common codebook $\mathbf{C} = \left[ \mathbf{c}_1, \ldots, \mathbf{c}_M \right] \in \mathbb{C}^{n\times M}$ with each element drawn i.i.d. according to $\mathcal{CN}(0,P')$.
    The total variation distance between the measures with and without power constraint can be bounded as
    \begin{align}
      p_{0,K_a} 
      & = K_a \mathbb{P}\left[ \chi^2(2n)  \geq \frac{2nP}{P'} \right] \\
      & \leq \min_{0\leq r < \frac{1}{2}} \left\{ K_a \exp\left\{-r\frac{2nP}{P'}\right\} \mathbb{E}\left[ \exp\left\{ r \chi^2(2n) \right\} \right] \right\} \label{eq_noCSI_noKa_p0_asymn1} \\
      & = \min_{0\leq r < \frac{1}{2}} \left\{ K_a \exp\left\{-r\frac{2nP}{P'} - n \ln (1-2r) \right\} \right\} \label{eq_noCSI_noKa_p0_asymn2}  \\
      & = K_a \exp\left\{-n \left( \frac{P}{P'} - 1 - \ln \frac{P}{P'} \right) \right\},\label{eq_noCSI_noKa_p0_asymn3}
    \end{align}
    where \eqref{eq_noCSI_noKa_p0_asymn1} follows from the Chernoff bound $\mathbb{P}\left[W>0\right] \leq \mathbb{E}\left[ \exp\left\{rW\right\} \right]$ for $r\geq 0$~\cite{elements_IT}; \eqref{eq_noCSI_noKa_p0_asymn2} follows from the moment-generating function 
    $\mathbb{E}\left[ \exp\left\{ t \chi^2(k) \right\} \right] = (1-2t)^{-k/2}$ for $t<\frac{1}{2}$;
    and \eqref{eq_noCSI_noKa_p0_asymn3} follows because $ r\frac{2nP}{P'} + n \ln (1-2r)$ is maximized in the case of $r = \frac{1}{2}\left( 1-\frac{P'}{P} \right)$.
    When $n \to \infty$ and $\frac{P'}{P}=c$ with the constant $c\in(0,1)$, we have $p_{0,K_a} \to 0$.

    The term $\frac{\binom{K_a}{2}}{M}$ in Corollary~\ref{Theorem_Ka_asymn} denotes an upper bound on the probability of existing message collision.
    In the following, we proceed to obtain $p_{K_a \to K'_a,1}$ and $p_{K_a \to K'_a,2}$ assuming all users share a common Gaussian codebook and there is no message collision among $K_a$ active users.
    Let the matrix $\mathbf{C}_{K_a} \in \mathbb{C}^{n\times K_a}$ include the transmitted codewords of $K_a$ active users. Following from Theorem~\ref{Theorem_Ka}, we have
    \begin{align}
      & p_{K'_a,1} \notag\\
      & = \!\mathbb{E} \! \left[ \min_{\rho\geq0} \exp \!\left\{  \rho nL \left( 1+\tilde{K}'_a  P^{\prime} \right) \right.\right. \notag\\
      & \;\;\;\;\; \;\;\;\;\; \;\;\;\;\; \;\;\;\;\; \;\;\;\;\;  \left.   -L \ln  \left|  \mathbf{I}_n + \rho \left( \mathbf{I}_n + \mathbf{C}_{K_a} \mathbf{C}_{K_a}^H \right) \right| \Big\} \right] \\
      & = \!\mathbb{E} \! \left[ \min_{\rho\geq0} \exp \!\left\{  \rho nL \!\left( 1+\tilde{K}'_a  P^{\prime} \right)
      - L(n-K_a)\ln(1+\rho) \right.\right. \notag\\
      & \;\;\;\;\; \;\;\;\;\; \;\;\;\;\; \;\;\;\;\; \;\;\;\;\;  \left. - L \ln \! \left| \mathbf{I}_{K_a} \!+\! \rho \left( \mathbf{I}_{K_a} \!+\! \mathbf{C}_{K_a}^H \mathbf{C}_{K_a} \right) \right| \Big\} \right] \!.
    \end{align}

    In the case of $\tilde{K}'_a < K_a$, $n\to\infty$, and $\rho n\to c_1$ with constant $c_1>0$, we have
    \begin{align}
      & p_{ K'_a,1} \notag\\
      & = \min_{\rho\geq0} \exp \!\left\{  \rho nL \left( 1+\tilde{K}'_a  P^{\prime} \right)
      - L(n-K_a)\ln(1+\rho) \right. \notag\\
      & \;\;\;\;\; \;\;\;\;\; \;\;\;\;\; \;\;\;\; - L K_a \ln  \left( 1+\rho(1\!+\! P' n) \right) \Big\} \!+ o(1) \label{eq_noCSI_noKa_p1_asymn1}\\
      & = \min_{\rho\geq0} \exp \left\{  \rho nL \left( 1+\tilde{K}'_a  P^{\prime} \right)
      - Ln\ln(1+\rho) \right. \notag\\
      & \;\;\;\;\; \;\;\;\;\; \;\;\;\;\; \;\;\;\;    - L K_a \ln  \left( 1+\rho P' n \right) \Big\}  + o(1) \label{eq_noCSI_noKa_p1_asymn2}\\
      & = \min_{\rho\geq0} \exp \left\{  \rho nL \tilde{K}'_a  P^{\prime} - L K_a \ln  \left( 1+\rho P' n \right) \right\}  + o(1) \label{eq_noCSI_noKa_p1_asymn3}\\
      & = \min_{\tilde{\rho}\geq0} \exp \left\{  \tilde{\rho} L \tilde{K}'_a - L K_a \ln  \left( 1+\tilde{\rho} \right) \right\}  + o(1) \label{eq_noCSI_noKa_p1_asymn4}\\
      & = \exp \left\{ K_a L \left(  1 - \frac{\tilde{K}'_a}{K_a} + \ln \frac{\tilde{K}'_a}{K_a} \right)  \right\}  + o(1),\label{eq_noCSI_noKa_p1_asymn5}
    \end{align}
    where \eqref{eq_noCSI_noKa_p1_asymn1} follows because the eigenvalues of $\frac{1}{P'} \mathbf{C}_{K_a}^H \mathbf{C}_{K_a}$ converge to $n+o(n)$ as $n\to\infty$~\cite{eigenvalue_wishart};
    \eqref{eq_noCSI_noKa_p1_asymn3} follows because $\rho n - n\ln(1+\rho) = \mathcal{O}(\rho^2 n) = o(1)$; \eqref{eq_noCSI_noKa_p1_asymn4} holds by setting $\tilde{\rho} = \rho P' n$; \eqref{eq_noCSI_noKa_p1_asymn5} follows because $\tilde{\rho} L \tilde{K}'_a - L K_a \ln  \left( 1+\tilde{\rho} \right)$ is minimized when $\tilde{\rho} = \frac{K_a}{\tilde{K}'_a}-1$ in the case of $\tilde{K}'_a < K_a$.
    When $\tilde{K}'_a \geq K_a$, we denote $p_{K_a\to K'_a,1} = 1$.

    Likewise, we can obtain $p_{ K'_a,2}$ under the assumption of $\tilde{K}'_a < K_a$ and $n\to\infty$.
    It concludes the proof of Corollary~\ref{Theorem_Ka_asymn}.

\section{Proof of Theorem~\ref{Theorem_achievability}} \label{Appendix_proof_achievability}

  In this appendix, we prove Theorem~\ref{Theorem_achievability} to establish an achievability bound on the minimum required energy-per-bit for the scenario in which the number $\mathrm{K}_a$ of active users is random and unknown.
  Specifically, we use a random coding scheme to generate a common codebook $\mathbf{C} = \left[ \mathbf{c}_{1}, \ldots, \mathbf{c}_{M} \right] \in \mathbb{C}^{n\times M}$ satisfing $\mathbb{E} [ \left\| \mathbf{c}_m \right\|_2^2  ] = nP'$ for $m \in [M]$ as introduced in the proof of Theorem~\ref{Theorem_Ka}.
  If user $k$ is active, it transmits $\mathbf{x}_{W_k} = \mathbf{c}_{W_k} 1 \big\{ \left\|\mathbf{c}_{W_k}\right\|_{2}^{2} \leq n P \big\}$. The transmitted messages of active users, i.e. ${\mathcal{W}} = \left\{W_{k}: k\in\mathcal{K}_a\right\}$, are sampled independently with replacement from $[M]$.
  To upper-bound the error probability, we perform three changes of measure:
  1)~the messages in $\mathcal{W}$ are sampled uniformly without replacement from $[M]$ and we have $\left|{\mathcal{W}}\right| = \mathrm{K}_a$ in this case;
  2)~the active user~$k$ transmits $\mathbf{x}_{W_k} = \mathbf{c}_{W_k}$;
  3)~there are at least $K_l$ and at most $K_u$ active users, i.e., $K_l \leq \mathrm{K}_a \leq K_u$, where $0 \leq K_l \leq K_u\leq M$.
  The total variation distance between the true measure and the new one is bounded by $p_0$ in~\eqref{eq_noCSI_p0}.

  The decoder first obtains an estimate of the number of active users as introduced in Section~\ref{Section:Ka}.
  Then, based on the MAP criterion, the decoder produces a set of decoded messages of size $\hat{K}_a$ belonging to an interval around the estimated value $K'_a$.
  That is, it is satisfied that $\hat{K}_a \in [K'_{a,l}, K'_{a,u}]$,
  where $K'_{a,l} = \max\left\{ K_l , K'_{a} -r' \right\}$, $K'_{a,u} = \min\left\{ K_u , K'_{a} +r' \right\}$, and $r'$ denotes a nonnegative integer referred to as the decoding radius.
  The per-user probability of misdetection can be upper-bounded as
  \begin{align}
    P_{\mathrm{MD}}
    & = \mathbb{E}  \left[ 1 \left[ \mathrm{K}_a > 0 \right]
            \cdot \frac{1}{\mathrm{K}_a} \sum_{k\in {\mathcal{K}_a}} \mathbb{P} \left[ W_{k} \notin \hat{\mathcal{W}}  \right] \right] \\
    & \leq \mathbb{E}  \left[ 1 \left[ \mathrm{K}_a > 0 \right]
            \cdot \frac{1}{\mathrm{K}_a}  \sum_{k\in {\mathcal{K}_a}} \mathbb{P}  \left[ W_{k} \notin \hat{\mathcal{W}}  \right] \right]_{\text{new}}   +  p_0 \label{eq_PUPE_MD_upper_noCSI_noKa1_r} \\
    & \leq   \sum_{K_a=\max\{K_l,1\}}^{K_u}  \!\!\!  P_{{\rm{K}}_a}(K_a)\!\!
    \sum_{K'_a=K_l}^{K_u}
    \sum_{t \in \mathcal{T}_{K'_a} }\!\!\!
    \frac{t+(K_a - K'_{a,u})^{+}}{K_a}  \notag\\
    & \;\;\;\;\; \cdot \mathbb{P} \left[  \mathcal{F}_{t} \cap \left\{ K_a \to K'_{a} \right\} \right]_{\text{new}} +  p_0. \label{eq_PUPE_MD_upper_noCSI_noKa_r_weight}
  \end{align}
  Here, the integer $t$ takes value in $\mathcal{T}_{K'_a}$ defined in~\eqref{eq_noCSI_noKa_estimate_Tset1} because the number of misdetected codewords, given by $t+(K_a-K'_{a,u})^{+}$, is lower-bounded by $(K_a-K'_{a,u})^{+}$ and upper-bounded by the total number $K_a$ of transmitted messages and by $M-K'_{a,l}$ since at least $K'_{a,l}$ messages are returned;
  $\mathcal{F}_{t} = \left\{ \sum_{k\in {\mathcal{K}_a}} 1 \left\{ W_k \notin \hat{\mathcal{W}} \right\} = t+(K_a-{K}'_{a,u})^{+} \right\}
  $ denotes the event that there are exactly $t+(K_a-K'_{a,u})^{+}$ misdetected codewords;
  $\left\{K_a \to K'_{a} \right\}$ denotes the event that the estimation of $K_a$ results in $K'_a$.
  Likewise, the per-user probability of false-alarm can be upper-bounded as
  \begin{align}
    P_{\mathrm{FA}}
    & = \mathbb{E}   \left[ 1 \left[ \hat{K}_a > 0 \right] \cdot  \frac{1}{ \hat{K}_a } \sum_{ i \in \hat{\mathcal{W}} } \mathbb{P} \left[ i \notin \mathcal{W}  \right]\right] \\
    & \leq    \sum_{K_a=K_l}^{K_u}  \!\!\!  P_{{\rm{K}}_a}(K_a) \!\!   \sum_{K'_a=K_l}^{K_u}
    \sum_{t \in \mathcal{T}_{K'_a} }
    \sum_{t' \in \mathcal{T}_{K'_a,t} }\!\!\!
        \frac{t' +  ( K'_{a,l}-K_a )^{+}}{ \hat{K}_a  } \notag\\
    & \;\;\;\;\; \cdot \mathbb{P}   \left[  \mathcal{F}_{t,t'}  \cap \left\{ K_a \to K'_{a} \right\} \right]_{\text{new}}
    +  p_0 ,\label{eq_PUPE_FA_upper_noCSI_noKa_r_weight}
  \end{align}
  where $\hat{K}_a$ denotes the number of detected codewords as given in~\eqref{eq_noCSI_noKa_estimate_Kahat};
  $\mathcal{F}_{t,t'}$ denotes the event that there are exactly $t+(K_a-K'_{a,u})^{+}$ misdetected codewords and $t' + ( K'_{a,l} - K_a )^{+}$ falsely alarmed codewords;
  the integer $t'$ takes value in $\mathcal{T}_{K'_a}$ defined in Theorem~\ref{Theorem_achievability} because:
  i)~$\hat{K}_a$ must be in $[K'_{a,l} : K'_{a,u} ]$;
  ii)~the number of falsely alarmed codewords is lower-bounded by $( K'_{a,l} - K_a )^{+}$ and upper-bounded by $M-K_a$;
  iii)~there exist falsely alarmed codewords only when $\hat{K}_a\geq 1$.

  Next, we omit the subscript ``new'' for the sake of brevity.
  The probability in the RHS of~\ref{eq_PUPE_MD_upper_noCSI_noKa_r_weight} can be bounded as
  \begin{align}
    & \mathbb{P} \left[  \mathcal{F}_{t} \cap \left\{ K_a \to K'_{a} \right\} \right] \notag\\
    & = \mathbb{P} \left[  \mathcal{F}_{t}
    \cap \left\{ \hat{K}_{a} \in [K'_{a,l} , K'_{a,u} ] \right\}
    \cap \left\{ K_a \to K'_{a} \right\} \right] \label{eq_PUPE_MD_FA_upper_noCSI_noKa1_r_weight} \\
    & \leq \min \left\{ \mathbb{P}  \left[  \mathcal{F}_{t}
    \cap \left\{ \hat{K}_{a}  \in  [K'_{a,l} , K'_{a,u} ] \right\} \right]
    , \mathbb{P} \left[ K_a  \to  K'_{a} \right] \right\} \label{eq_PUPE_MD_FA_upper_noCSI_noKa2_r_weight}\\
    & \leq \min \!\left\{ \sum_{t' \in \bar{\mathcal{T}}_{K'_a,t} } \!\!\!  \mathbb{P} \! \left[  \mathcal{F}_{t,t'}
    \left|  \hat{K}_{a} \! \in \! [K'_{a,l},K'_{a,u} ] \right.\right]
    , \mathbb{P} \left[ K_a  \to  K'_{a}  \right] \right\} \!. \label{eq_PUPE_MD_FA_upper_noCSI_noKa3_r_weight}
  \end{align}
  Here, $\bar{\mathcal{T}}_{K'_a,t}$ is defined in \eqref{eq_noCSI_noKa_estimate_Tset2bar}, which is obtained similar to $\mathcal{T}_{K'_a,t}$ with the difference that the number $\hat{K}_a$ of detected codewords can be $0$;
  \eqref{eq_PUPE_MD_FA_upper_noCSI_noKa1_r_weight} follows because the event $K_a \to K'_{a}$ implies that $ \hat{K}_{a} \in [K'_{a,l} , K'_{a,u} ]$~\cite{noKa};
  \eqref{eq_PUPE_MD_FA_upper_noCSI_noKa2_r_weight} follows from the fact that the joint probability is upper-bounded by each of the individual probabilities.
  Similarly, the probability in the RHS of~\eqref{eq_PUPE_FA_upper_noCSI_noKa_r_weight} can be bounded as
  \begin{align}
    &  \mathbb{P} \left[  \mathcal{F}_{t,t'} \cap \left\{ K_a \to K'_{a} \right\} \right] \notag\\
    &  \leq \min \left\{ \mathbb{P} \left[  \mathcal{F}_{t,t'}
    \left| \hat{K}_{a} \in [ K'_{a,l} , K'_{a,u} ] \right.\right]
    , \mathbb{P} \left[ K_a \to K'_{a} \right] \right\} . \label{eq_PUPE_MD_FA_upper_noCSI_noKa_r}
  \end{align}
  The upper bound on $\mathbb{P} \left[ K_a \to K'_{a} \right]$ is given in Theorem~\ref{Theorem_Ka}.
  Then, we proceed to bound the probability $\mathbb{P} \left[  \mathcal{F}_{t,t'}
    \left| \hat{K}_{a} \in [ K'_{a,l} , K'_{a,u} ] \right.\right]$.

  We use the MAP decoder to obtain the estimated set $\hat{\mathcal{W}}$ of the transmitted messages.
  The decoder outputs
  \begin{equation}
    \hat{\mathcal{W}} = \left\{\; \emph{f}_{\text{en}}^{-1}\left( \hat{\mathbf{c}} \right): \hat{\mathbf{c}} \in \hat{\mathcal{C}}  \right\},
  \end{equation}
  \begin{equation}\label{eq:decoderoutput_noCSI}
    \hat{\mathcal{C}}
    =\arg \min_{ \hat{\mathcal{C}} \subset \{ \mathbf{c}_1, \ldots, \mathbf{c}_M \}: | \hat{\mathcal{C}} | \in [ K'_{a,l} , K'_{a,u} ] }
    g\;\! ( \hat{\boldsymbol{\Gamma}} ) ,
  \end{equation}
  where both the sets $\hat{\mathcal{W}}$ and $\hat{\mathcal{C}}$ are of size $\hat{K}_a$,
  the matrix $\hat{\boldsymbol{\Gamma}} = \operatorname{diag} \left\{ \hat{\boldsymbol{\gamma} } \right\}\in\{0,1\}^{M\times M}$ with $[\hat{\boldsymbol{\gamma} }]_i=1$ if $\mathbf{c}_i \in \hat{\mathcal{C}}$,
  and the MAP decoding metric $g ( \hat{\boldsymbol{\Gamma}} )$ is given by
  \begin{align}
    g ( \hat{\boldsymbol{\Gamma}} )
    = & L \ln \big| \mathbf{I}_n + \mathbf{C}\hat{\boldsymbol{\Gamma}}\mathbf{C}^H \big|
    +  \operatorname{tr}  \left(  \mathbf{Y}^{H}
    \big( \mathbf{I}_n  +  \mathbf{C}\hat{\boldsymbol{\Gamma}}\mathbf{C}^H \big)^{-1}   \mathbf{Y}  \right) \notag\\
    & -  \ln  \left(  \frac{P_{{\rm{K}}_a} (\hat{K}_a) }{ \binom{M}{\hat{K}_a} } \right)   .\label{eq:g_noCSI}
  \end{align}

  Let the set $\mathcal{W}_1 \subset \mathcal{W}$ of size $t+(K_a-K'_{a,u})^{+}$ denote the set of misdecoded messages.
  The set $\mathcal{W}_1$ can be divided into two subsets $\mathcal{W}_{1,1}$ and $\mathcal{W}_{1,2}$ of size $(K_a-K'_{a,u})^{+}$ and $t$, respectively.
  Let the set $\mathcal{W}_2 \subset [M]\backslash \mathcal{W}$ of size $t'+(K'_{a,l}-K_a)^{+}$ denote the set of false-alarm codewords.
  Let $\mathcal{W}_{2,1}$ denote an arbitrary subset of $\mathcal{W}_2$ of size $(K'_{a,l}-K_a)^{+}$.
  For the sake of simplicity, we rewrite ``$\cup_{ \mathcal{W}_{1} \subset \mathcal{W}, \;\! \left| \mathcal{W}_{1} \right| = t+(K_a-K'_{a,u})^{+} }  $'' to ``$\cup_{\mathcal{W}_{1}}$''
  and ``$\cup_{{\mathcal{W}_{2} \subset [M] \backslash \mathcal{W},\;\! \left| \mathcal{W}_{2} \right| = t'+(K'_{a,l}-K_a)^{+} } }$'' to ``$\cup_{\mathcal{W}_2}$''; similarly for $\sum$ and $\cap$.
  Then, we have
  \begin{align}
      & \mathbb{P}  \left[  \mathcal{F}_{t,t'} \left|  \hat{K}_a \in [K'_{a,l} , K'_{a,u} ] \right.\right] \notag\\
      & \leq \mathbb{P}   \left[ \left. \mathcal{G}_e
      \right| \hat{K}_a \in [{K}'_{a,l} ,{K}'_{a,u} ] \right] \label{eq:proof_noCSI_noKa_pftt1} \\
      & \leq \! \min_{ 0 \leq \omega \leq 1 , \nu\geq0 }\!
      \left\{  \mathbb{P} \! \left[ \left. \mathcal{G}_e
      \cap \mathcal{G}_{\omega,\nu}
      \right| \hat{K}_a  \!\in\!  [{K}'_{a,l} ,{K}'_{a,u} ] \right]
       +  \mathbb{P}   \left[ \mathcal{G}_{\omega,\nu}^c  \right]
      \right\}\! ,  \label{eq:proof_noCSI_noKa_pftt}
  \end{align}
  where $\mathcal{G}_e = {\cup}_{\mathcal{W}_{1}}  {\cup}_{\mathcal{W}_{2}}
       \left\{ g  \left( {\boldsymbol{\Gamma}}_{ \mathcal{W}  \backslash \mathcal{W}_1 \cup \mathcal{W}_2 }   \right)
       \leq  g  \left( \boldsymbol{\Gamma}_{ \mathcal{W} \backslash \mathcal{W}_{1,1} \cup \mathcal{W}_{2,1} } \right)  \right\}$, and \eqref{eq:proof_noCSI_noKa_pftt1} holds because we treat all ties in $\mathcal{G}_e$ as errors.
  According to the properly selected region around the linear combination of the transmitted signals given in~\cite[Eq.~(14)]{TIT}, we define the event
  \begin{equation}
    \mathcal{G}_{\omega,\nu} = \bigcap_{\mathcal{W}_1}  \left\{ g\left( \boldsymbol{\Gamma}_{\mathcal{W}} \right) \leq \omega g\left( {\boldsymbol{\Gamma}}_{ \mathcal{W} \backslash \mathcal{W}_1 } \right) + nL\nu \right\}.
  \end{equation}
  Then, \eqref{eq:proof_noCSI_noKa_pftt} follows from Fano's bounding technique $\mathbb{P}\left[A\right] \leq \mathbb{P}\left[A\cap B\right] +\mathbb{P}\left[ B^c \right]$ given in~\cite{1961}.

  The first probability in the RHS of~\eqref{eq:proof_noCSI_noKa_pftt} can be bounded as
    \begin{align}
      & \mathbb{P} \!\left[ \left. \mathcal{G}_e
      \cap \mathcal{G}_{\omega,\nu}
      \right| \hat{K}_a \in [{K}'_{a,l} ,{K}'_{a,u} ] \right] \notag\\
      & \leq \! \sum_{ \mathcal{W}_1, \mathcal{W}_2} \mathbb{E} \! \left[ \mathbb{P}  \Big[ \!\left\{ g \left( {\boldsymbol{\Gamma}}_{ \mathcal{W}  \backslash \mathcal{W}_1 \cup \mathcal{W}_2 }   \right)
      \leq g \left( \boldsymbol{\Gamma}_{ \mathcal{W} \backslash \mathcal{W}_{1,1} \cup \mathcal{W}_{2,1} } \right)  \right\} \right.  \notag\\
      & \;\;\;\;\;
      \left.\left.\left.
      \cap \left\{ g\!\left( \boldsymbol{\Gamma}_{\mathcal{W}} \right) \!\leq\! \omega g\!\left( {\boldsymbol{\Gamma}}_{ \mathcal{W} \backslash \mathcal{W}_1 } \right) \!+\! nL\nu \right\}
      \right| \!\mathbf{C} , \hat{K}_a \!\in\! [{K}'_{a,l} ,\!{K}'_{a,u} ] \right] \right] \label{eq_noCSI_q1t_union} \\
      & \leq \! C_{K'_a,t,t'}
      \mathbb{E} \!\left[ \min_{ {0\leq u,0\leq r} } 
      \!\;\!\!\mathbb{E}_{\mathbf{H},\mathbf{Z}} \!\Big[ \;\!\! \exp \!\left\{ rnL\nu \!+\!
      u g \!\left( \boldsymbol{\Gamma}_{ \mathcal{W} \backslash \mathcal{W}_{1,1} \cup \mathcal{W}_{2,1} } \right)
       \right. \right. \notag\\
      & \;\;\;\;\;   - u g\!\left( {\boldsymbol{\Gamma}}_{ \mathcal{W}  \backslash \mathcal{W}_1 \cup \mathcal{W}_2 }   \right)
      - r g \!\left(\boldsymbol{\Gamma}_{ \mathcal{W} } \right) \notag\\
      & \;\;\;\;\;  \! \;\!\!
      \left.\left.\left.
      + \;\! r\omega \;\! g\!\left(  {\boldsymbol{\Gamma}}_{ \mathcal{W} \backslash \mathcal{W}_1 } \right) \right\}       \right| \!\mathbf{C} , \hat{K}_a \in [{K}'_{a,l} ,{K}'_{a,u} ] \right] \bigg],  \label{eq_noCSI_q1t_chernoff}
    \end{align}
  where the term $C_{K'_a,t,t'} = \binom{K_a}{ t+(K_a-K'_{a,u})^{+} } \binom{M-K_a}{t'+(K'_{a,l}-K_a)^{+}}$;
  \eqref{eq_noCSI_q1t_union} follows from the union bound;
  \eqref{eq_noCSI_q1t_chernoff} follows by applying the Chernoff bound $\mathbb{P}\left[\{Z \geq 0\} \cap \{W \geq 0\}\right] \leq \mathbb{E}\left[\exp\left\{u Z + r W\right\}\right]$ for $u,r \geq 0$~\cite[Section 3.2.4]{goodregion}, to the conditional probability in~\eqref{eq_noCSI_q1t_union}.
  Define $\mathbf{F}$, $\mathbf{F}_1$, $\mathbf{F}'$, $\mathbf{F}''$, and $b_{u,r}$ as in Theorem~\ref{Theorem_achievability}.
  Denote the conditional expectation over $\mathbf{H}$ and $\mathbf{Z}$ in the RHS of~\eqref{eq_noCSI_q1t_chernoff} as $E_1$, which can be written as
  \begin{align}
      E_1 \!
      & =  \exp\! \left\{
      u L\! \ln\!\left| \mathbf{F}'' \right|
      \!-\! u L\! \ln\!\left| \mathbf{F}' \right|
      \!-\! r L\! \ln\!\left| \mathbf{F} \right|
      \!+\! r\omega L\! \ln\!\left| \mathbf{F}_1 \right|
      \!+\! b_{u,r}   \right\}    \notag\\
      & \;\;\;  \cdot \mathbb{E}_{ \mathbf{H},\mathbf{Z} } \! \left[ \left. \exp \! \left\{
      \operatorname{tr}\!\left( \mathbf{Y}^H \! \left(
      u \left(\mathbf{F}''\right)^{-1}
      \!-\! u \left( \mathbf{F}' \right)^{-1}
      \!-\! r \mathbf{F}^{-1} \right.\right.\right.\right.\right.\notag\\
      &  \;\;\;\;\; \;\;\;\;\; \;\;\;\;\;  \left.\left.\left.\left.\left. + r\omega \mathbf{F}_{1}^{-1} \right)\mathbf{Y} \right)
      \right\} \right| \mathbf{C} , \hat{K}_a \!\in\! [{K}'_{a,l} ,{K}'_{a,u} ] \right]  \\
      & = \exp  \left\{ L \left(
      u \ln \!\left|\mathbf{F}''\right|
      \!-\! u \ln \!\left| {\mathbf{F}'} \right|
      - r \ln \!\left|\mathbf{F}\right|
      + r\omega \ln \!\left| \mathbf{F}_{1} \right| \right.\right. \notag\\
      & \;\;\;\;\; \;\;\;\;\; \;\;\;\;\;\; \left.\left. - \ln \!\left| \mathbf{B} \right| \right)  + b_{u,r}
      \right\}   \label{eq_noCSI_q1t_exp_y}.
  \end{align}
  Here, \eqref{eq_noCSI_q1t_exp_y} follows from Lemma \ref{expectation_bound} by taking the expectation over $\mathbf{H}$ and $\mathbf{Z}$ provided that the minimum eigenvalue of $\mathbf{B}$ satisfies  $\lambda_{\min}\left(\mathbf{B}\right) > 0$, where the matrix $\mathbf{B}$ is given by
      \begin{equation}
        \mathbf{B} = (1+r) \mathbf{I}_n
        - u \left( \mathbf{F}'' \right)^{-1} \mathbf{F}
        + u \left( \mathbf{F}' \right)^{-1} \mathbf{F}
        - r\omega \mathbf{F}_{ 1}^{-1} \mathbf{F}.
      \end{equation}
  Substituting \eqref{eq_noCSI_q1t_exp_y} into \eqref{eq_noCSI_q1t_chernoff}, we can obtain an upper bound on $\mathbb{P}  \left[ \left. \mathcal{G}_e
      \cap \mathcal{G}_{\omega,\nu}
      \right| \hat{K}_a \in [{K}'_{a,l} ,{K}'_{a,u} ] \right]$, which is denoted as $q_{1,K'_a,t,t'} \left(\omega,\nu\right)$.

  Define the event $\mathcal{G}_\delta =\bigcap_{\mathcal{W}_1} \left\{ \sum_{i=1}^{n}  {\chi_i^2(2L)}  \leq  2nL(1+\delta) \right\}$ for $\delta\geq0$.
  In the case of $t+(K_a-K'_{a,u})^{+} > 0$, we can bound the second probability $\mathbb{P}  \left[ \mathcal{G}_{\omega,\nu}^c  \right]$ in the RHS of~\eqref{eq:proof_noCSI_noKa_pftt} as
  \begin{equation} \label{eq:noCSI_q2t_goodregion}
    \mathbb{P} \left[ \mathcal{G}_{\omega,\nu}^c  \right]
    \leq \min_{\delta\geq 0}
    \left\{  \mathbb{P} \left[ \mathcal{G}_{\omega,\nu}^c \cap \mathcal{G}_\delta  \right]
    + \mathbb{P} \left[ \mathcal{G}_\delta^c \right]  \right\}.
  \end{equation}
  Here, $\mathbb{P} \left[ \mathcal{G}_\delta^c \right] = \binom{K_a}{t+(K_a-K'_{a,u})^{+}} \left( 1 - \frac{\gamma\left( nL, nL \left( 1+\delta \right)\right)}{\Gamma\left( nL \right)} \right)$
  and $\mathbb{P} \left[  \mathcal{G}_{\omega,\nu}^c \cap \mathcal{G}_\delta \right]$ is bounded as
    \begin{align}
      & \mathbb{P}  \left[ \mathcal{G}_{\omega,\nu}^c \cap \mathcal{G}_{\delta} \right] \notag\\
      & = \mathbb{P} \! \left[ \bigcup_{\mathcal{W}_1} \left\{ \sum_{l=1}^{L} \left( {\mathbf{y}}_l^H \left( \mathbf{F}^{-1}  - \omega  \mathbf{F}_1^{-1} \right) {\mathbf{y}}_l \right) > \bar{b} \right\} \cap \mathcal{G}_\delta \right]\label{eq:q2t_y} \\
      & = \!\mathbb{E} \!\left[  \mathbb{P} \!\left[ \bigcup_{\mathcal{W}_1} \left. \!\left\{ \sum_{i=1}^{n} \!\left(1\!-\!\omega\!-\! \omega \lambda_i\right) \!\frac{\chi_i^2(2L)}{2}
      \!>\!   \bar{b} \right\}
      \cap  \mathcal{G}_\delta \right| \mathbf{C} , \mathcal{W}
      \right] \right] \label{eq:q2t_tildey}\\
      & \leq \! \sum_{\mathcal{W}_1}  \mathbb{E} \!\left[  \mathbb{P} \! \left[ \left.  \sum_{i=1}^{m} \! \frac{\lambda_i \chi_i^2(2L)}{2}  \!<\!  \frac{nL(1\!+\!\delta)(1\!-\!\omega)\!-\! \bar{b} }{\omega}  \right| \!\mathbf{C} , \mathcal{W}
      \right] \right] \label{eq:noCSI_q3t_cap} \\
      & \leq \!\binom{K_a}{t+(K_a-K'_{a,u})^{+}} \mathbb{E} \! \left[
      \frac{\gamma \!\left( Lm, \frac{nL(1+\delta)(1-\omega)- \bar{b} }{\omega \prod_{i=1}^{m} \lambda_i^{ {1}/{m}}  } \right)}{\Gamma\left( Lm \right)}  \right] \!, \label{eq:noCSI_q2t_prod}
    \end{align}
  where the term $\bar{b} = \omega L \ln \left|\mathbf{F}_1\right| - L \ln \left|\mathbf{F}\right| + n L \nu - \omega b_1 + b $;
  and $\lambda_1, \ldots, \lambda_n$ denote the eigenvalues of the matrix $\mathbf{F}_1^{-1} \mathbf{C}\boldsymbol{\Gamma}_{\mathcal{W}_1}\mathbf{C}^H$ of rank $m = \min \left\{ n, t+(K_a-K'_{a,u})^{+} \right\}$ in decreasing order.
  Here, \eqref{eq:q2t_tildey} holds because $\mathbf{y}_{l} =  \mathbf{F}^{\frac{1}{2}} \tilde{\mathbf{y}}_{l} \stackrel{ \rm{i.i.d.} }{\sim} \mathcal{CN} \left( \mathbf{0}, \mathbf{F} \right)$ conditioned on $\mathbf{C}$ and $\mathcal{W}$ with $\tilde{\mathbf{y}}_{l} \stackrel{ \rm{i.i.d.} }{\sim} \mathcal{CN} \left( \mathbf{0}, \mathbf{I}_n \right)$ for $l\in[L]$;
  \eqref{eq:noCSI_q3t_cap} follows from the union bound and the inequality $\mathbb{P}\left[\{Z \geq 0\} \cap \{W \geq 0\}\right] \leq \mathbb{P}\left[ u Z + r W \geq 0 \right]$ for $u,r\geq 0$;
  and \eqref{eq:noCSI_q2t_prod} follows by applying Lemma~\ref{lemma_chisumprod} shown below.
  Together with~\eqref{eq:noCSI_q2t_goodregion}, we obtain an upper bound on $\mathbb{P}  \left[ \mathcal{G}_{\omega,\nu}^c \right]$ in the case of $t+(K_a-K'_{a,u})^{+} > 0$, which is denoted as $q_{2,K'_a,t}\left(\omega,\nu\right)$.
  \begin{Lemma}[\cite{chisumprod}] \label{lemma_chisumprod}
    Assume $x_i \stackrel{\text {i.i.d.}}{\sim} \chi^{2}\left(m\right)$ for $i=1,2,\ldots,s$ and $\tilde{x} \sim \chi^{2}\left(sm\right)$.
    Let the constant $\gamma_j>0$. Then, for every constant~$c$, we have
    \begin{equation}
      \mathbb{P} \left[{\sum}_{j=1}^{s} \gamma_{j} x_{j} <c\right] \leq \mathbb{P} \left[ {\prod}_{j=1}^{s} \gamma_{j}^{\frac{1}{s}} \tilde{x} <c\right].
    \end{equation}
  \end{Lemma}

  When $t+(K_a-K'_{a,u})^{+} = 0$, we have
  \begin{align}
    \mathbb{P}  \left[ \mathcal{G}_{\omega,\nu}^c \right]
    & = \mathbb{P} \left[ g\left(\boldsymbol{\Gamma}_{\mathcal{W}} \right) >  \frac{nL\nu}{1-\omega} \right] \\
    & = \mathbb{E} \! \left[ 1 - \frac{\gamma\left( nL, \frac{nL\nu}{1-\omega} - L \ln \left| \mathbf{F} \right|  +  b \right)}{\Gamma\left( nL \right)}  \right] , \label{eq:proof_noCSI_noKa_pft_t0_rbig}
  \end{align}
  where the constant $b$ is given in Theorem~\ref{Theorem_achievability}.
  The RHS of \eqref{eq:proof_noCSI_noKa_pft_t0_rbig} is denoted as $q_{2,K'_a,t,0}$ as in~\eqref{eq_noCSI_noKa_estimation_q2t0}.
  This concludes the proof of Theorem~\ref{Theorem_achievability}.


\section{Proof of Theorem~\ref{Theorem_converse_single_noKa}} \label{Appendix_proof_converse_single_noKa}

%
%
%
%

  We assume there are $K$ users and each of them becomes active with probability $p_a$ independently.
  The converse bound for the scenario with knowledge of the activities of $K-1$ users and the transmitted codewords and channel coefficients of active users among them, is also converse in the case without this knowledge.
  If the remaining user is active, it equiprobably selects a message $W \in [M]$, and the transmitted codeword satisfies the maximum power constraint in~\eqref{eq:power_constraint}.
  Let $\mathcal{F} \subset \mathbb{C}^n$ be a set of channel inputs satisfying~\eqref{eq:power_constraint}.
  Denote the set of decoded message as $\hat{\mathcal{W}}$ of size $|\hat{\mathcal{W}}| \leq K$.
  The requirements on the per-user probabilities of misdetection and false-alarm are given by
    \begin{equation} \label{eqR:MD}
      P_{e,{\rm{MD}}}  =   \frac{p_a}{ M }  \sum_{m \in [M]}  \mathbb{P} \big[ m  \notin  \hat{\mathcal{W}}  | W  =  m \big]   \leq  p_a \epsilon_2  =  \epsilon_{\rm{MD}},
    \end{equation}
    \begin{equation} \label{eqR:FA2}
      P_{e,{\rm{FA}}}  \leq  (1 - p_a) \mathbb{P} \big[ \hat{\mathcal{W}} \neq \emptyset | W  =  0 \big] \leq (1-p_a) \epsilon_1 = \epsilon_{\rm{FA}}.
    \end{equation}

  In the case of $|\hat{\mathcal{W}}|\leq 1$, an upper bound on the number of codewords that are compatible with the requirements on the probabilities of false-alarm, misdetection, and inclusive error is provided in~\cite[Theorem~2]{On_joint}.
  By enlarging the list size from $|\hat{\mathcal{W}}|\leq 1$ to $|\hat{\mathcal{W}}|\leq K$, changing the error requirement in~\cite{On_joint} to~\eqref{eqR:MD} and~\eqref{eqR:FA2}, and considering MIMO fading channels, we can obtain the following proposition.
  \begin{prop} \label{R_Theorem_AWGN_singleUE_conv}
    Consider the single-user setup with active probability $p_a$.
    Let $Q_{ Y^{n\times L} }$ be an arbitrary distribution on $\mathcal{Y}^{n\times L}$.
    The $(n,M,\epsilon_{\rm{MD}},\epsilon_{\rm{FA}},P)$ code satisfies
    \begin{equation} \label{eqR:beta_conv_AWGN_singleUE_1}
        \frac{M}{K} \leq\!\! \sup_{  { {P_{X^{n}}\!: \mathbf{x} \in \mathcal{F} , \epsilon_{1}\!,\epsilon_{2} \in [0,1]} } }
        \frac{ 1 - \beta_{1-\epsilon_{1}}  \left( P_{ Y^{n\times L} | X^{n} = \mathbf{0} } , Q_{ Y^{n\times L} } \right) }{ \beta_{ 1-\epsilon_{2} }
        \!\left( P_{ X^{n} } P_{Y^{n\times L}|X^{n}} , P_{X^{n}} Q_{Y^{n\times L}} \right)  } ,
      \end{equation}
    where
      \begin{equation}
        p_a \epsilon_{2} \leq \epsilon_{\rm{MD}} ,
      \end{equation}
      \begin{equation}
        (1-p_a) \epsilon_{1} \leq \epsilon_{\rm{FA}} .
      \end{equation}
    \begin{IEEEproof}
      Let $P_{X^{n}}$ be the distribution on $X^{n}$ induced by the encoder when the messages are uniform. Let $P_{Y^{n\times L}}$ be the output distribution. By assumption, we have
      \begin{equation}\label{eqR:proof_epsilon1}
        P_{ Y^{n\times L} \mid X^{n}=\bf{0} } \big[ \hat{\mathcal{W}} = \emptyset \big] \geq 1-\epsilon_1.
      \end{equation}
      Let $Z = 1\big[ \big\{ W\in\hat{\mathcal{W}} \big\} \cap \big\{ \hat{\mathcal{W}} \neq \emptyset \big\} \big]$. We have
      \begin{equation}\label{eqR:proof_PXY}
        P_{X^{n}, Y^{n\times L}}[Z=1]=P_{X^{n}, Y^{n\times L}}[ W\in\hat{\mathcal{W}} ] \geq 1-\epsilon_2,
      \end{equation}
      \begin{align}\label{eqR:proof_PXQY}
        & P_{X^{n}} Q_{Y^{n\times L}}[Z=1] \notag\\
        & = \sum_{W\in[M]} \frac{1}{M}
        \sum_{\hat{\mathcal{W}} \in \binom{[M]}{\leq K} } Q (\hat{\mathcal{W}}) 1\left[ \left\{ W \in \hat{\mathcal{W}} \right\} \cap \left\{ \hat{\mathcal{W}} \neq \emptyset \right\} \right] \notag\\
        & = \frac{1}{M} \sum_{\hat{\mathcal{W}} \in \binom{[M]}{\leq K} } Q (\hat{\mathcal{W}})
        \; |\hat{\mathcal{W}}| \notag\\
        & \leq \frac{K  Q_{Y^{n\times L}} \left[ \hat{\mathcal{W}} \neq \emptyset \right] }{M}.
      \end{align}
      It then follows that
      \begin{equation}\label{eqR:proof_beta_epsilon2}
        \beta_{1-\epsilon_2}  \left( P_{X^{n}, Y^{n\times L}}, P_{X^{n}} Q_{Y^{n\times L}} \right)
         \leq  \frac{ K}{M}  Q_{Y^{n\times L}}   \big[ \hat{\mathcal{W}} \neq \emptyset \big] .
      \end{equation}
      The term $Q_{Y^{n\times L}} \left[ \hat{\mathcal{W}} \neq \emptyset \right]$ can be further bounded as
      \begin{align}
        Q_{Y^{n\times L}}   \big[ \hat{\mathcal{W}} \neq \emptyset \big]
        & =  1-Q_{Y^{n\times L}} \big[ \hat{\mathcal{W}} = \emptyset \big] \\
        & \leq  1-\beta_{1-\epsilon_1} \left( P_{ Y^{n\times L} \mid X^{n}=\bf{0} }, Q_{Y^{n\times L}} \right) .\label{eqR:proof_beta_epsilon2_Q}
      \end{align}
      Substituting \eqref{eqR:proof_beta_epsilon2_Q} into \eqref{eqR:proof_beta_epsilon2}, \eqref{eqR:beta_conv_AWGN_singleUE_1} can be obtained.
    \end{IEEEproof}
  \end{prop}

  Next, we follow similar lines as in~\cite[Appendix~I]{TIT} to loosen Proposition~\ref{R_Theorem_AWGN_singleUE_conv} and obtain an easy-to-evaluate bound in~\eqref{eqR:beta_conv_AWGN_singleUE_1e}, which completes the proof of Theorem~\ref{Theorem_converse_single_noKa}.


%
\section{Proof of Theorem~\ref{Theorem_converse_noCSI_Gaussian_cKa}} \label{Appendix_proof_converse_noCSI_Gaussian_noKa}

  When there are $K_a$ active users, we assume w.l.o.g. that the active user set is $\mathcal{K}_a = [K_a]$.
  Let $\mathbf{X}\in\mathbb{C}^{n\times M}$ be the codebook matrix. Denote $\bar{\mathbf{X}}_{K_aM} = \left[\mathbf{X},\mathbf{X},\ldots,\mathbf{X}\right] \in \mathbb{C}^{n\times K_aM}$ and $\bar{\mathbf{X}} = \operatorname{diag} \left\{\bar{\mathbf{X}}_{K_aM}, \bar{\mathbf{X}}_{K_aM}, \ldots, \bar{\mathbf{X}}_{K_aM} \right\} \in \mathbb{C}^{nL\times K_aML}$.
  Let $\bar{\mathbf{H}}_{l}$ be a $K_a M  \times K_a M$ block diagonal matrix, whose block $k$ is a diagonal $M \times M$ matrix with diagonal entries equal to ${h}_{k,l}$.
  Let $\bar{\mathbf{H}} = \left[\bar{\mathbf{H}}_{1}, \ldots , \bar{\mathbf{H}}_{L} \right]^T$.
  The vector $\bar{\boldsymbol{\beta}} \in \left\{ 0,1 \right\}^{K_aM}$ has $K_a$ blocks, whose block $k$ denoted as $\bar{\boldsymbol{\beta}}_k$ is of size $M$ and includes one $1$.
  Assume $\bar{\mathbf{z}} \sim \mathcal{CN} ({\bf{0}} , \mathbf{I}_{nL} )$. We have
  \begin{equation}
    \bar{\mathbf{y}} = \bar{\mathbf{X}}  \bar{\mathbf{H}}  \bar{\boldsymbol{\beta}} + \bar{\mathbf{z}} \in \mathbb{C}^{nL\times 1}.
  \end{equation}

  For $k\in[K_a]$, define $\mathcal{M}_k = 1 [ W_k \notin \hat{\mathcal{W}}] $,
  where $\hat{\mathcal{W}}$ denotes the decoded list of size $| \hat{\mathcal{W}} | = \hat{K}_a \leq K$.
  Let $P_{e,k} = \mathbb{P} [ \mathcal{M}_k = 1 ]$,
  $\bar{P}_{e, K_a} = \frac{1}{K_a}\sum_{k\in [K_a]} P_{e,k}$ if $K_a \geq 1$, and $\bar{P}_{e, K_a} = 0$ if $K_a =0$.
  We have $P_{e} = \sum_{K_a \geq 1} P_{{\rm K}_a}(K_a) \bar{P}_{e,K_a} \leq \epsilon_{\rm MD}$.
  For any user $k\in[K_a]$, the standard Fano inequality gives that
  \begin{align}
    &H_2( W_k | \bar{\mathbf{X}} , {\rm{K}}_a = K_a )
    - H_2\left( \mathcal{M}_k | \bar{\mathbf{X}} , {\rm{K}}_a = K_a \right) \notag\\
    & - H_2 ( W_k | \mathcal{M}_k , \hat{\mathcal{W}} , \bar{\mathbf{X}} , {\rm{K}}_a = K_a ) \notag\\
    & \leq I_2  (   \mathcal{W}_{k};\hat{\mathcal{W}} \left| \bar{\mathbf{X}} , {\rm{K}}_a = K_a \right. ) , \label{eq:fano_noKa}
  \end{align}
  where the first three entropies are given by $H_2( W_k | \bar{\mathbf{X}} , {\rm{K}}_a = K_a ) = J$,
  $ H_2( \mathcal{M}_k | \bar{\mathbf{X}} , {\rm{K}}_a = K_a )
    = h_2(P_{e,k})$, and
  \begin{align}
    & H_2 ( W_k | \mathcal{M}_k , \hat{\mathcal{W}} , \bar{\mathbf{X}} , {\rm{K}}_a  =  K_a ) \notag\\
    & \leq (1 - P_{e,k})   \log_2  \hat{K}_a  +  P_{e,k} J \label{eq:fano_noCSI_entropy1} \\
    & \leq \log_2 \hat{K}_a + P_{e,k} J. \label{eq:fano_noCSI_entropy}
  \end{align}

  Then, by taking the average over $k\in[K_a]$ and $K_a \in S \subset \mathcal{K}$ on both sides of~\eqref{eq:fano_noKa}, and applying Jensen's inequality and $\sum_{k\in[K_a]}  I_2 (  \mathcal{W}_{k};\hat{\mathcal{W}}  \left| \bar{\mathbf{X}} , {\rm{K}}_a  =  K_a \right.  )  \leq  I_2 \big(  \mathcal{W}_{[K_a]};\hat{\mathcal{W}}  \left| \bar{\mathbf{X}} , {\rm{K}}_a  =  K_a  \right.\big)$, we can obtain
  \begin{align}
    & \left( \mathbb{P}[ {\rm K}_a \in S ] - \sum_{K_a \in S} P_{{\rm K}_a}(K_a) \bar{P}_{e,K_a} \right) J \notag\\
    & - \sum_{K_a \in S} P_{{\rm K}_a}(K_a) h_2 \left( \bar{P}_{e,K_a} \right)
    - \sum_{K_a \in S} P_{{\rm K}_a}(K_a) \log_2  \hat{K}_a \notag\\
    & \leq
    \sum_{K_a \in S} P_{{\rm K}_a}(K_a)  \frac{1}{K_a} I_2 \big(  \mathcal{W}_{[K_a]};\hat{\mathcal{W}} \left| \bar{\mathbf{X}} , {\rm{K}}_a  =  K_a \right.  \big) . \label{eq:fano_noCSI_Ka2}
  \end{align}
  It is satisfied that
  \begin{align}
    & \sum_{K_a \in S} P_{{\rm K}_a}(K_a) \bar{P}_{e,K_a} J
    + \sum_{K_a \in S} P_{{\rm K}_a}(K_a) h_2\left( \bar{P}_{e,K_a} \right) \notag\\
    & \leq P_e J + h_2(P_e) \label{eq:fano_noCSI_Ka2_left1} \\
    & \leq \epsilon_{\rm MD} J + h_2(\epsilon_{\rm MD}), \label{eq:fano_noCSI_Ka2_left2}
  \end{align}
  where \eqref{eq:fano_noCSI_Ka2_left1} follows because $S \subset \mathcal{K}$ and by applying Jensen's inequality for $h_2(\cdot)$ and \eqref{eq:fano_noCSI_Ka2_left2} holds under the condition that $P_e \leq \epsilon_{\rm MD} \leq \frac{M}{1+M}$.

  The mutual information in~\eqref{eq:fano_noCSI_Ka2} can be upper-bounded as~\cite[Eq.~(149)]{finite_payloads_fading}
    \begin{align}
      I_2 \big(  \mathcal{W}_{[K_a]};\hat{\mathcal{W}} \left| \bar{\mathbf{X}} , {\rm{K}}_a = K_a \right.  \big)
      & \leq
      I_2  \left( \bar{\mathbf{H}}\bar{\boldsymbol{\beta}}; \bar{\mathbf{y}} | \;\!  \bar{\mathbf{X}} , {\rm{K}}_a =  K_a  \right) \notag\\
      & \!\!\!\!\! \!\!\!\!\! \!\!\!\!\! \!\!\!\!\!-  I_2  \left( \bar{\mathbf{H}}\bar{\boldsymbol{\beta}}; \bar{\mathbf{y}} | \;\! \bar{\boldsymbol{\beta}}, \bar{\mathbf{X}} , {\rm{K}}_a   =  K_a  \right) .\label{eq_conv_fano_nocsi}
    \end{align}

    Next, we focus on the two terms on the RHS of~\eqref{eq_conv_fano_nocsi}.
    We have
    \begin{align}
      & I_2\left(\left. \bar{\mathbf{H}}\bar{\boldsymbol{\beta}};\bar{\mathbf{y}} \right| \bar{\mathbf{X}} = \bar{\mathbf{X}}^r , {\rm{K}}_a = K_a \right) \notag\\
      & =  I_2\left( \left. \bar{\mathbf{H}}\bar{\boldsymbol{\beta}}; \bar{\mathbf{X}}^r\bar{\mathbf{H}}\bar{\boldsymbol{\beta}}+ \bar{\mathbf{z}}
      \right| {\rm{K}}_a = K_a \right)  \\
      & \leq \sup_{\mathbf{u}} I_2\left( \left. \mathbf{u}; \bar{\mathbf{X}}^r\mathbf{u}+\bar{\mathbf{z}} \right| {\rm{K}}_a = K_a \right)   \label{eq_conv_inf_sup}\\
      &  = L \log_2 \left| \mathbf{I}_{n } + \frac{K_a}{M} \mathbf{X}^{r} \left( \mathbf{X}^{r}\right)^{H} \right|,\label{eq_conv_inf_sup3}
    \end{align}
    where $ \mathbf{X}^r$ is a realization of $\mathbf{X}$ and $\bar{\mathbf{X}}^r$ is a realization of $\bar{\mathbf{X}}$.
    The supremum in \eqref{eq_conv_inf_sup} is over
    $\mathbf{u}$ with $\mathbb{E}  \left[\mathbf{u} \right]  =  \mathbf{0}$ and $\mathbb{E}  \left[\mathbf{u} \mathbf{u}^{H} \right]  =  \mathbb{E}  \left[ \left(\bar{\mathbf{H}}\bar{\boldsymbol{\beta}} \right) \left(\bar{\mathbf{H}}\bar{\boldsymbol{\beta}} \right)^{ H} \right]  =  \frac{1}{M}\mathbf{I}_{K_aML}$.
    The supremum is achieved when $\mathbf{u}  \sim  \mathcal{CN}  \left( \mathbf{0}, \frac{1}{M}\mathbf{I}_{K_aML} \right)$ \cite{sparsity_pattern}, which implies \eqref{eq_conv_inf_sup3}.
    Then, we have
    \begin{align}
      & I_2 \left( \left. \bar{\mathbf{H}}\bar{\boldsymbol{\beta}}; \bar{\mathbf{y} } \right| \bar{\mathbf{X}} , {\rm{K}}_a = K_a \right) \notag\\
      & \leq L \mathbb{E}\!  \left[  \left. \log_2 \left| \mathbf{I}_{n} + \frac{K_a}{M} \mathbf{X} \mathbf{X}^{ H}  \right|  \; \right|   {\rm{K}}_a = K_a \right] \\
      & \leq nL \log_2 \left( 1 + K_aP \right), \label{eq_conv_inf_term1}
    \end{align}
    where \eqref{eq_conv_inf_term1} follows from the concavity of the $\log_2 \left|\cdot\right|$ function and Jensen's inequality under the assumption that the entries of codebooks are i.i.d. with mean $0$ and variance $P$.
  Following from (43) and (44) in~\cite{letter}, we have
    \begin{align}
      & I  \left( \bar{\mathbf{H}}\bar{\boldsymbol{\beta}};\bar{\mathbf{y}} | \bar{\boldsymbol{\beta}}, \bar{\mathbf{X}}, {\rm{K}}_a = K_a \right) \notag\\
      & \geq  L \mathbb{E}  \left[ \log_2 \left| \mathbf{I}_{n} + \mathbf{X}
      \sum_{k=1}^{K_a} \operatorname{diag}(\bar{\boldsymbol{\beta}}_k)
      {\mathbf{X}}^{H} \right|  \right] \label{eq_conv_inf_term2_yury} \\
      & \geq \left( 1- \frac{1}{M} {\binom{K_a}{2}} \right) L \mathbb{E} \left[  \log_2  \left| \mathbf{I}_{n} + \mathbf{X}_{K_a}  \mathbf{X}_{K_a}^{ H} \right|  \right] \label{eq_conv_inf_term2},
    \end{align}
    where 
    $\mathbf{X}_{K_a}$ includes $K_a$ different codewords.

    Substituting \eqref{eq:fano_noCSI_Ka2_left2}, \eqref{eq_conv_fano_nocsi}, \eqref{eq_conv_inf_term1}, and \eqref{eq_conv_inf_term2} into \eqref{eq:fano_noCSI_Ka2}, we can obtain~\eqref{P_tot_conv_noCSI} in Theorem~\ref{Theorem_converse_noCSI_Gaussian_cKa}.
    Moreover, we can obtain \eqref{P_tot_conv_noCSI_FA} considering that at least $\max \{0,\hat{K}_a-K_a\}$ messages are false-alarmed when the size of the decoded list is $\hat{K}_a$.

\clearpage
\IEEEtriggeratref{11}

\end{document}